%% file: main.tex
\documentclass[twocolumn,twoside]{Jornadas}
\usepackage[utf8]{inputenc}
\usepackage{listings}
\usepackage{adjustbox}
\usepackage{graphicx}
\usepackage{color}
\usepackage{caption}
\usepackage{comment}
\usepackage{amsmath}
\usepackage{amsfonts}
\usepackage{amssymb}
\usepackage[caption=false,font=footnotesize]{subfig}
\usepackage{hyperref}
\usepackage{url}
\usepackage{cite}
\usepackage{graphicx}
\usepackage{float}
\usepackage{placeins}
\usepackage{balance}
\usepackage{url}
\usepackage{algorithm2e}
\DeclareMathVersion{normal2}
\mathversion{normal2}
\usepackage{mathrsfs}
\mathversion{normal}
\usepackage{bm}

\newcommand\blfootnote[1]{%
  \begingroup
  \renewcommand\thefootnote{}\footnote{#1}%
  \addtocounter{footnote}{-1}%
  \endgroup
}

\newtheorem{conj1}{Conjecture}
\newtheorem{corollary}{Corollary}[conj1]

\definecolor{gray97}{gray}{.97}
\definecolor{gray75}{gray}{.75}
\definecolor{gray45}{gray}{.45}

\def\BibTeX{{\rm B\kern-.05em{\sc i\kern-.025em b}\kern-.08em
    T\kern-.1667em\lower.7ex\hbox{E}\kern-.125emX}}

\graphicspath{{.}{./Figuras/}}

\begin{document}
\raggedbottom

\title{Improving the Engine of Society}

\author{Stephen Casey\thanks{NASA Langley Research Center; \tt{smc.contact.1@gmail.com}}}

\markboth{}{}
\pagestyle{plain} 
\thispagestyle{empty} 
\setcounter{section}{-1}
\setcounter{secnumdepth}{4}

\addtocontents{toc}{\setcounter{tocdepth}{2}}

\twocolumn[
  \begin{@twocolumnfalse}
    \maketitle
    \thispagestyle{empty}
    \tableofcontents
  \end{@twocolumnfalse}
]

\blfootnote{$^1$NASA Langley Research Center;\\ \tt{smc.contact.1@gmail.com}}
\newpage
\clearpage
 
 \twocolumn[
  \begin{@twocolumnfalse}
    \input{01-Intro.tex}
    \vspace{2pt}
  \end{@twocolumnfalse}
  ]
  

\input{02-Chapter1.tex}
\input{chaos01-Intro}
\input{chaos02-Theory}
\input{chaos03-Methods}
\input{chaos04-Results}
\input{chaos05-Conclusions}
\input{04-Chapter3.tex}

\input{05-Chapter4.tex}
\input{06-Conclusions.tex}
 
\nocite{*}
\bibliographystyle{unsrt}
\bibliography{biblio}

\end{document}

%% file: 01-Intro.tex
\section{Introduction}
\PARstart{T}{his} paper is organized into four chapters.  The first chapter observes how decoupling labor from ownership of the output produced by that labor is equivalent to building an engine that is incentivized to generate the maximum energy output, which leads to unstoppable climate change and compromised quality of life for many individuals.  The second chapter discusses the organization of complex systems into hierarchical objects and functions, and proposes an improved working definition for the information entropy contained in complex systems.  Chapter 3 redesigns the engine from Chapter 1 into a system optimized to maximize Complex Information Entropy (CIE) rather than energy expenditure, which leads to improvements in the climate, regrowth of environmental ecosystems, minimization of useless labor, and maximization of the well-being of participants.  Chapter 4 examines climate change specifically, and introduces a possible solution in the form of a digital twin with an entropy-based fitness function.
\vspace{8pt}
\section{Chapter 1\\Money as Potential Energy:\\The Thermodynamic Cause of the Problems in Modern Capitalism}

%% file: 02-Chapter1.tex
\PARstart{P}{revious} work in the fields of econophysics and thermoeconomics has attempted to find analogies between physical and economic processes.  Considering the way money is used today, it is easy to see the parallels between money and potential energy.  Money can be exchanged for goods or services which take energy to perform or produce.  On a larger scale, if the government directs money toward particular fields in engineering or science, or particular sectors of the economy, accelerated development will occur in those areas.  A worker who earns a salary is given money in exchange for their labor, which can be viewed as a conversion of kinetic energy - labor - to potential energy - their salary.  In every case, the metaphor is a useful lens through which to view financial activity.

However, what happens if the metaphor is removed?  If wealth is regarded as a form of energy, what sort of thermodynamic process is a capitalist economy?    An analysis of this type reveals the economy to be equivalent to an engine, which, due to the engine's settings, is incentivized to 1. operate as inefficiently as possible and 2. generate as much energy as possible.  Income inequality and environmental damage follow as natural and inevitable consequences.

\begin{figure}[b]
    \includegraphics[width=1\linewidth]{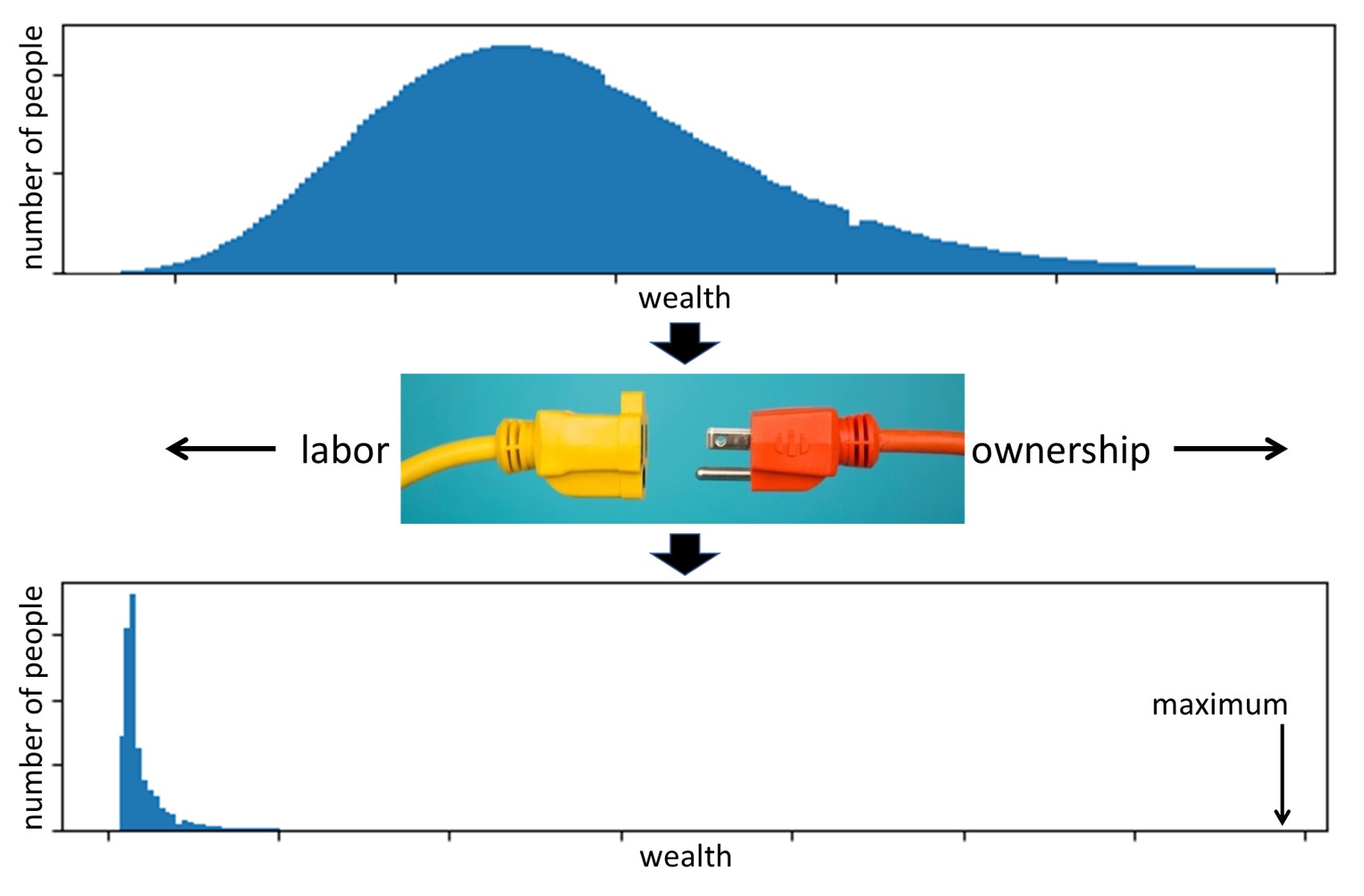}
    \caption{Decoupling labor from ownership squeezes the gamma distribution against the y axis and produces a long tail.}
    \label{fig:img53}
\end{figure}

This chapter begins an investigation into the fundamental thermodynamics that underlie the complexity of our financial systems.
\vspace{2.0ex plus .5ex minus .2ex}
\subsection{An Incomplete List of Problems}

\begin{figure*}
    \centering
    \includegraphics[width=1\textwidth]{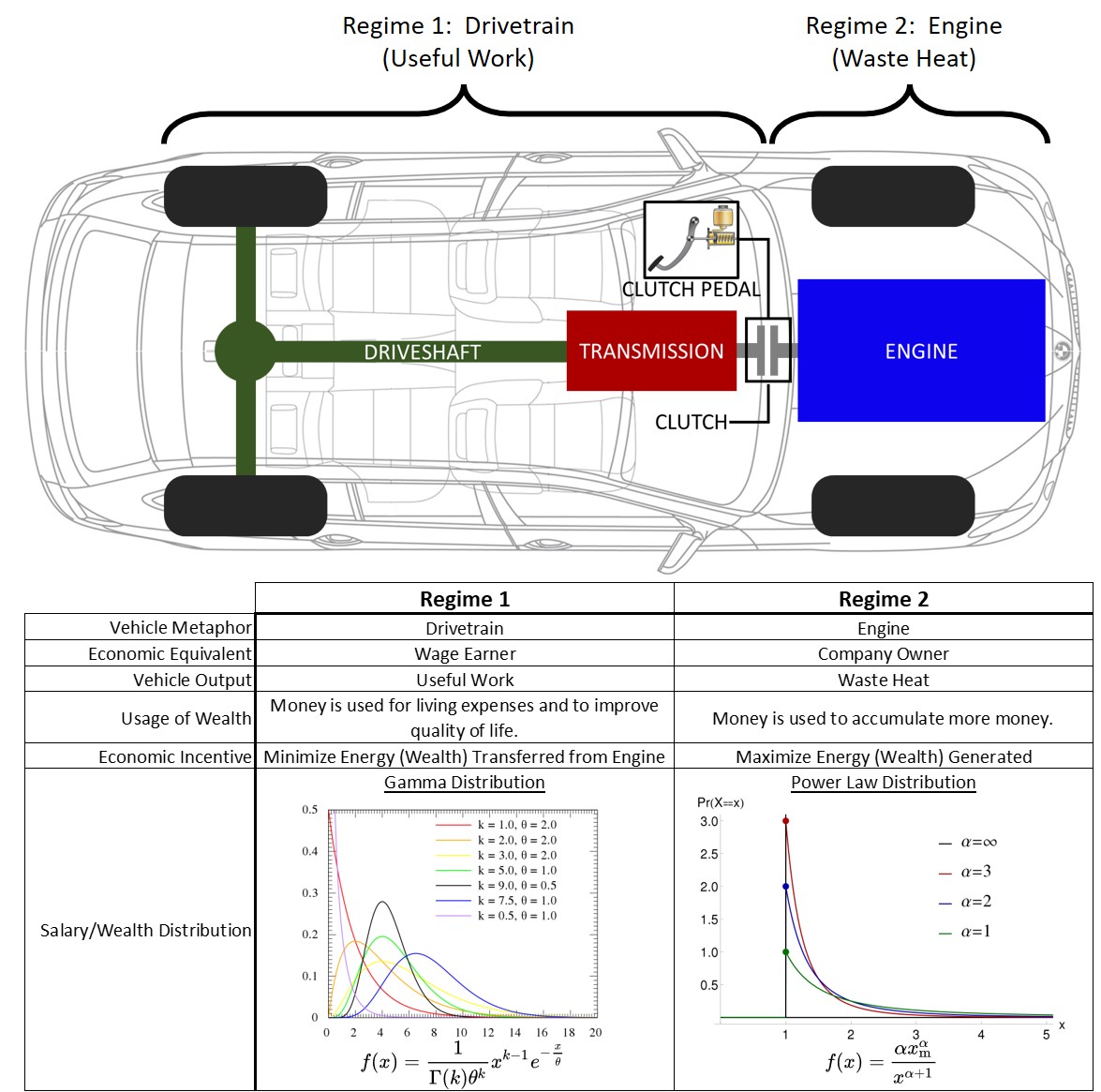}
    \caption{Across a wide variety of contexts and geographic locations, detailed analyses of income and wealth distributions are seen to follow this two-regime pattern.  Image credit:  \cite{gamma}\cite{powerlaw}}
    \label{fig:img1}
\end{figure*}

What follows is an incomplete list of problems present in modern capitalist economies.

\begin{itemize}
\item	The climate is being pushed to catastrophic failure.  The combined worldwide efforts of policymakers and activists to correct the problem have been ineffective.
\item	The need for people to work during the COVID-19 pandemic resulted in many avoidable deaths.
\item	High levels of income inequality result in high levels of poverty in ostensibly wealthy countries.
\item	Worldwide, hundreds of millions of people live in extreme poverty \cite{wb}.
\item	Some low-income countries are dependent on foreign aid.
\item	Geopolitical conflicts are often connected to economic factors.
\item	International corporations have increased the homogeneity of world cultures.
\item	Individual anomalies, such as a housing bubble, can collapse an entire economy. 
\item	A high percentage of jobs in wealthy countries are bullshit jobs, which refer to meaningless or unnecessary wage labor which the worker must pretend has a purpose \cite{graeber2018bullshit}.
\item	Cubicle farms exist.
\item	Some jobs exist that would be better not done, including the production of junk mail, robocalling, telemarketing, and scamming.
\item	Financial sector jobs do not produce anything.
\item   Some industries, such as insurance, banking, and brokering, exist only to move money around.
\item	Emotionally-manipulative advertising is omnipresent and inescapable.
\item	Relationships between individuals are mediated and complicated by the exchange of wealth.
\item	There is a neverending concern about finances throughout a person’s entire life.
\item	Some individuals develop the mindset that the pursuit of wealth is life’s top priority.
\item	Hustle culture and the gig economy have grown as a consequence of companies hiring independent contractors rather than employees.
\item	The phrase “it’s just business” is used to justify behavior that would be unethical in other contexts.
\item	Organizations that do not generate a profit must rely on donations.
\item	Charitable causes are subject to the whims of the wealthy or media attention.
\item	Socially beneficial projects are often not undertaken because they are not profitable.
\item	Innovations and creative activities are opposed rather than supported.  
\item	Activities conducted for personal growth are treated as luxuries rather than inherent goods.  Drudgery is treated as an inherent good rather than a necessary evil.
\item	Unemployment, overwork, underpay, and job dissatisfaction lead to unhappiness, suffering, neurosis, anxiety, and problems in interpersonal relationships.
\end{itemize}

\subsection{The Economy as an Engine}


\begin{figure*}
    \centering
    \includegraphics[width=1\textwidth]{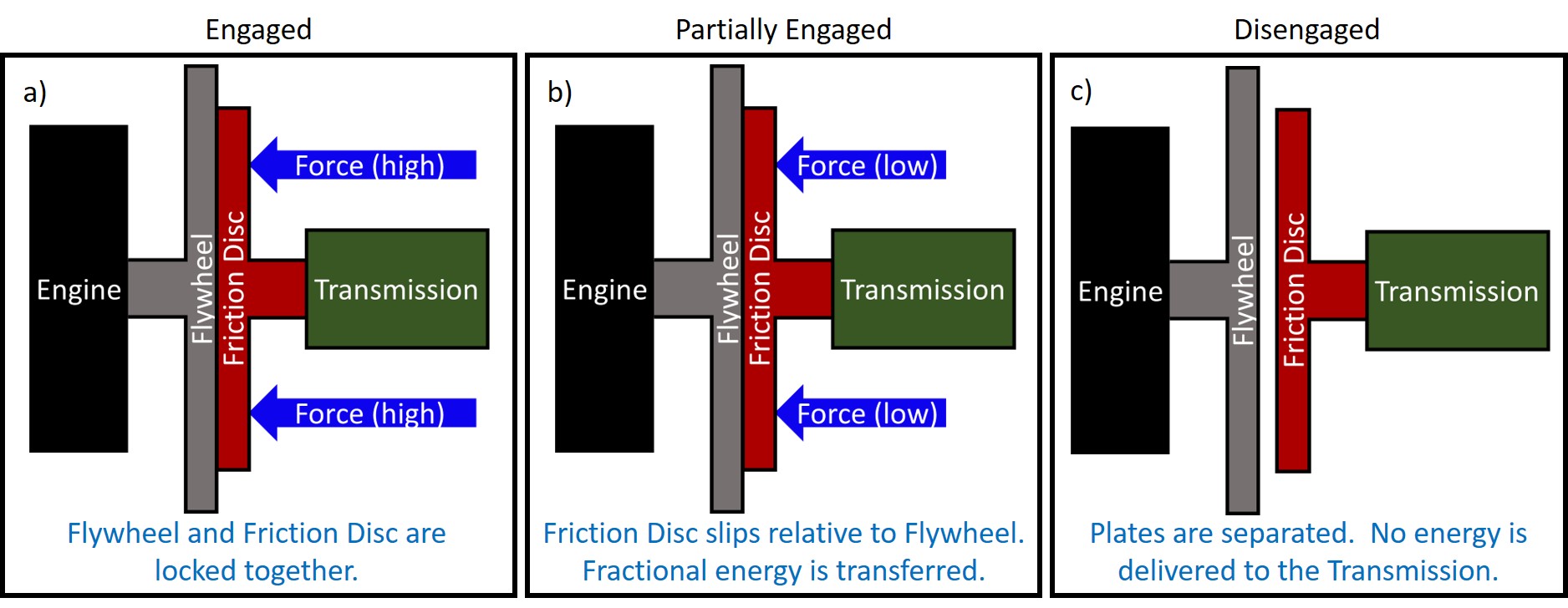}
    \caption{Panel c) depicts a scenario in which all employee salaries are zero, which would make a functioning society impossible.  Panel a) depicts a plausible scenario that is undesirable to capitalists.  Panel b) shows the configuration used in modern society, in which wealth transfer between rich and poor is minimized while energy output is maximized.}
    \label{fig:img2}
\end{figure*}

In order to perform this analysis, it's first necessary to determine the different forms of work and energy exchange present in the system.  In the modern incarnation of capitalism, the workers in a company are paid to produce goods or services, which can be sold for more than the workers are paid, and the surplus is kept by the company’s owners.  Labor is decoupled from ownership.  A diverse array of historical figures have noticed problems with this this arrangement.  In “Wages of Labor”, Karl Marx notes how the separation of labor from capital results in the demand for individuals to work longer hours, a monotonically increasing division and specialization of labor, and a constant need for more individuals to enter the workforce \cite{marx2009economic}.  Two centuries earlier, in his Second Treatise on Government \cite{locke2002second}, John Locke described the Labor Theory of Property whereby the fundamental concept of property itself arises by virtue of the labor an individual applies to a medium, and an object becomes the property of a person by virtue of the labor that person gives to the object\footnote{This is true both in a physical and psychological sense.  A person produces a creation in the physical world by combining his energy with raw materials, while also producing an inner representation of this object in his mind based on interactions with the physical object.   The individual is therefore an integral part of the physical object in the sense that the object would not exist without his actions, while he simultaneously feels ownership of the outer object due to the psychic energy he has directed toward his inner object.  This multi-level concurrence lends support to Locke’s theory.}
.  Classical economic theory describes how the opposing forces in salary negotiations will mathematically tend toward an equilibrium under ideal conditions.  This formulation shares similarities to many engineering problems.

The decoupling of labor from ownership clearly results in many interesting philosophic and mathematical questions.  Just as clearly, this decoupling has caused a great deal of anger, resentment, and conflict among owners and workers.  However, these symptoms seem to be emergent properties of a more fundamental cause.  We will now investigate the thermodynamics of this system.

One may imagine the economy as a car engine, capable of producing some amount of energy output.  Business owners are motivated to generate as much wealth as possible, which is equivalent to redlining the engine.  The incentive is therefore to press the gas petal to the floor at all times.

In a manual transmission drivetrain, a clutch connects the rotary motion of the engine to the transmission, propeller shaft, differential, wheels, and tires, as shown in Figure \ref{fig:img1}.  The clutch is responsible for transmitting the energy produced by the engine to the vehicle’s drivetrain.  One may imagine company workers as receiving the energy transmitted to the drivetrain, and company owners as receiving the energy produced by the engine.  The incentive therefore is to keep the engine revolutions as high as possible, while at the same time, keeping the clutch minimally engaged and transmitting as little energy as possible to the drivetrain.  The incentive to constantly redline the engine is likely the reason for the unstoppable progression of climate change.  This is discussed further in Chapter 3.

The clutch petal is operated by company owners, who determine how much wealth to transfer to workers.  If the clutch is fully engaged, as shown in Figure \ref{fig:img2}a, the engine and drivetrain are locked together, and all of the engine’s mechanical output is translated into useful work.  This is undesirable for the owners, as they accumulate less energy (wealth) in the form of waste heat.  However, if the clutch is fully disengaged, as shown in Figure \ref{fig:img2}c, the workers will not have enough energy (wealth) to afford food and shelter.  Therefore, the solution is to partially engage the clutch such that the flywheel slips against the friction disc and transfers the minimum amount of energy to the drivetrain, as shown in Figure \ref{fig:img2}b.  Therefore the system can be divided into two regimes:  Regime 1 is may be considered the Drivetrain Regime, Wage-Earner Regime, or Useful-Work Regime, while Regime 2 is similarly equivalent to the Engine Regime, Company-Owner Regime, or Waste-Heat Regime.  Across a wide variety of contexts and geographic locations, detailed analyses of income and wealth distributions are seen to follow this two-regime pattern.  The low end of the spectrum tends to follow a gamma distribution, while the high end exhibits a power law distribution \cite{Chatterjee_2007}.  As illustrated in Figure \ref{fig:img53}, the effect of disconnecting labor from ownership produces this two-regime distribution, which effectively squeezes the gamma distribution against $x=0$ as tightly as possible.
\subsection{Recoupling Labor to Ownership}

The rest of this chapter is dedicated to examining what would happen in a hypothetical company if labor and ownership were reconnected.  This is equivalent to fully coupling the engine to the drivetrain in Figure \ref{fig:img1}.  The employees in the hypothetical company begin the simulation with salary distribution similar to that of Amazon.com.

We will begin with a few basic assumptions.

\begin{figure}[t]
    \includegraphics[width=\linewidth]{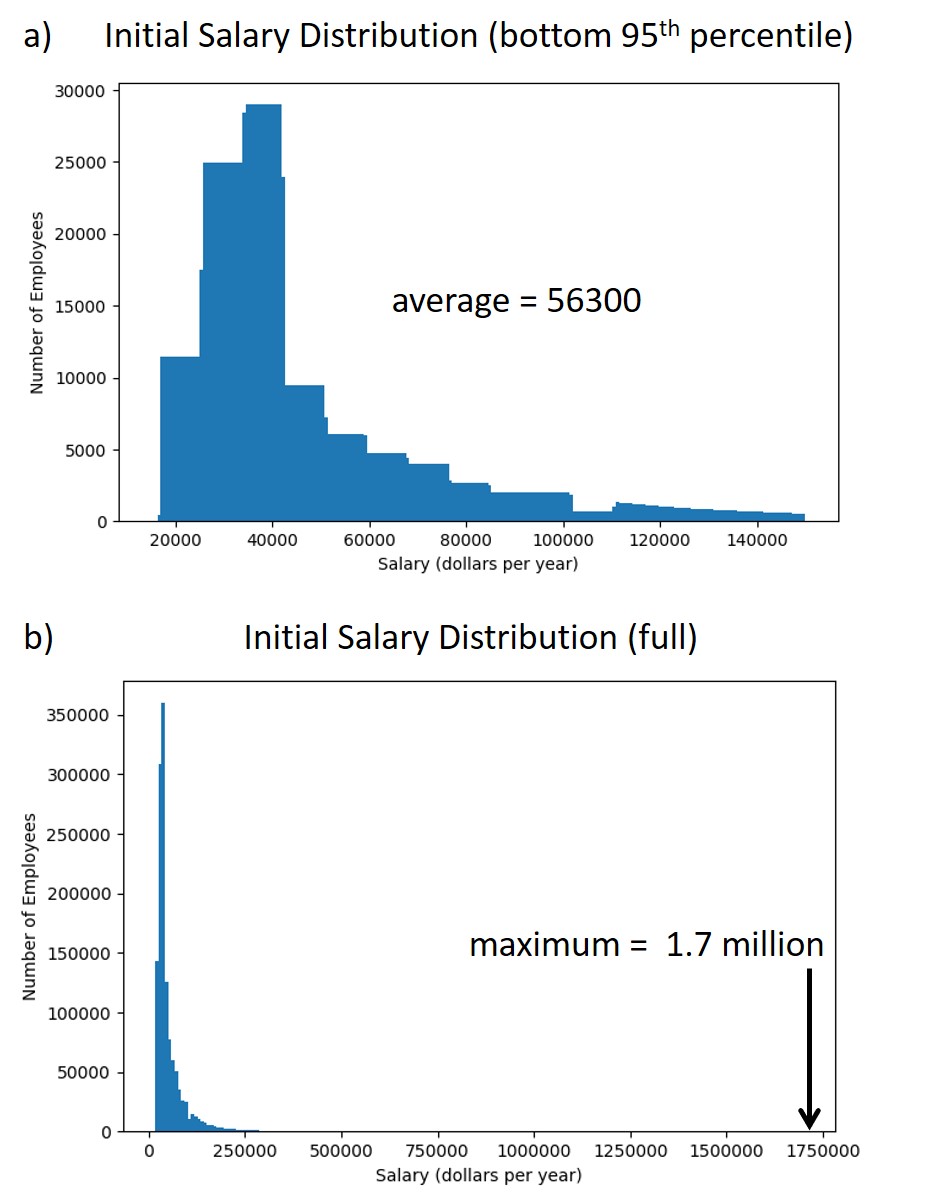}
    \caption{Initial company salary distribution}
    \label{fig:img3}
\end{figure}

\begin{enumerate}
   \item The total number of employees is assumed to be constant at 1,335,000, comparable to Amazon in 2020 \cite{statista}.
    
   \item A person is paid according to how much their labor contributes to a company.  For example, the work of a person who earns \$100,000 per year is hypothetically twice as important to a company’s operations as the work of a person who earns \$50,000 per year.  To define this concept precisely, salary is assumed to be a measurement of the information entropy of the function performed by each worker.  This is discussed in later chapters.
   
   \item The distribution of information entropy is assumed to follow a gamma distribution.
   \begin{equation}
        f(x) \sim \frac{x^{z-1}}{\theta^z}e^{-\frac{x}{\theta}}
   \end{equation}
   The values for $z$ and $\theta$ are assumed to be 11 and 5116, respectively.  A gamma distribution with $z=11$ is equivalent to a Maxwell-Boltzmann distribution with 11 dimensions.  The value of the thermal parallel for $\theta = 5116$ is unclear.  In a Maxwell-Boltzmann distribution, $\theta=kT$, where $T$ is the system temperature and $k$ is the Boltzmann constant, equal to $1.380649*10^{-23}$ Joules per Kelvin.  Conceptually, $k$ is a measurement of the amount of energy contained in a unit of temperature or entropy, which in this case has units of $\textrm{Joules} / \textrm{Shannon}$ or $\textrm{Joules} / (\textrm{dollars}/\textrm{year})$.  The conversion rate between Shannons and dollars/year has not yet been established.
   
   \item Comparable to a thermal system, the salary distribution relaxes to equilibrium exponentially. For an individual with starting salary $S_0$ and equilibrium salary $S_e$:
   \begin{equation}
        S(t) = S_0 - (S_0 - S_e)e^{-\frac{t}{\tau}} .
   \end{equation}
   The relaxation time constant $\tau$ is assumed to be 0.75 years.
   
   \item A person owns a fraction of a company equal to their lifetime earnings divided by the sum of the lifetime earnings of all employees.
   \begin{multline}
        \textrm{fraction of company owned by employee }n = \\ \frac{\textrm{total salary earned by employee }n}{\sum \textrm{total salary earned by all employees}}
    \end{multline}
   
    \item The total value of a company is considered to be its market capitalization, which is the total value of a publicly traded company's outstanding common shares owned by stockholders.  The market capitalization is assumed to be constant at \$1.46 trillion \cite{marketcap}.

\end{enumerate}

\begin{figure}
    \includegraphics[width=\linewidth]{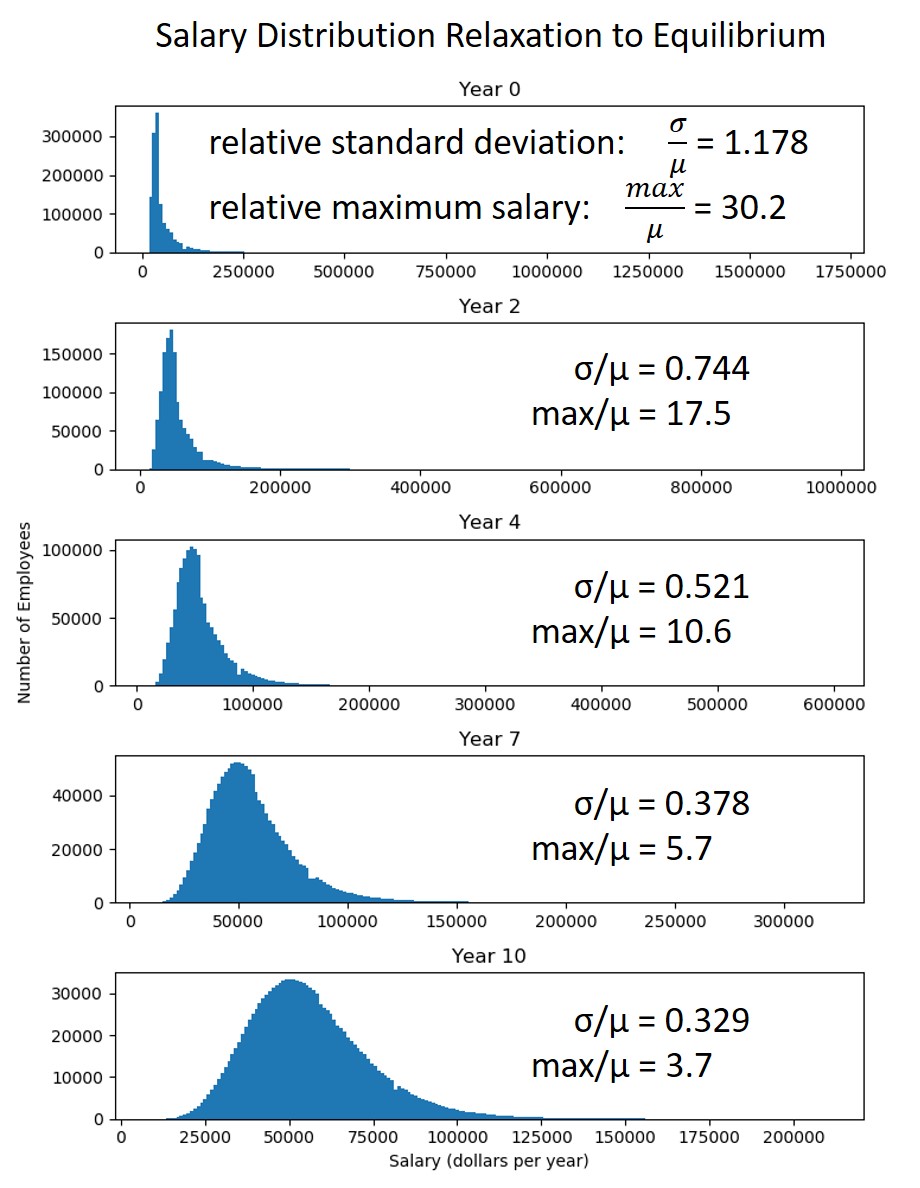}
    \caption{The employee salary distribution becomes more equitable as the system relaxes to equilibrium.}
    \label{fig:img4}
\end{figure}

Our hypothetical company begins with a low-end salary distribution equal to that for Amazon in July 2022 \cite{zip}.  The salaries above \$110,500 are assumed to follow a power law distribution.  The CEO's salary is \$1.7 million, based on Amazon's 2020 proxy statement stating that Jeff Bezos, Amazon's CEO, received a total compensation of \$1.7 million in 2020 \cite{marketwatch}.  The starting distributions for the bottom 95th percentile as well as the entire company are shown in Figures \ref{fig:img3}a and b.

\begin{figure}[t]
    \includegraphics[width=\linewidth]{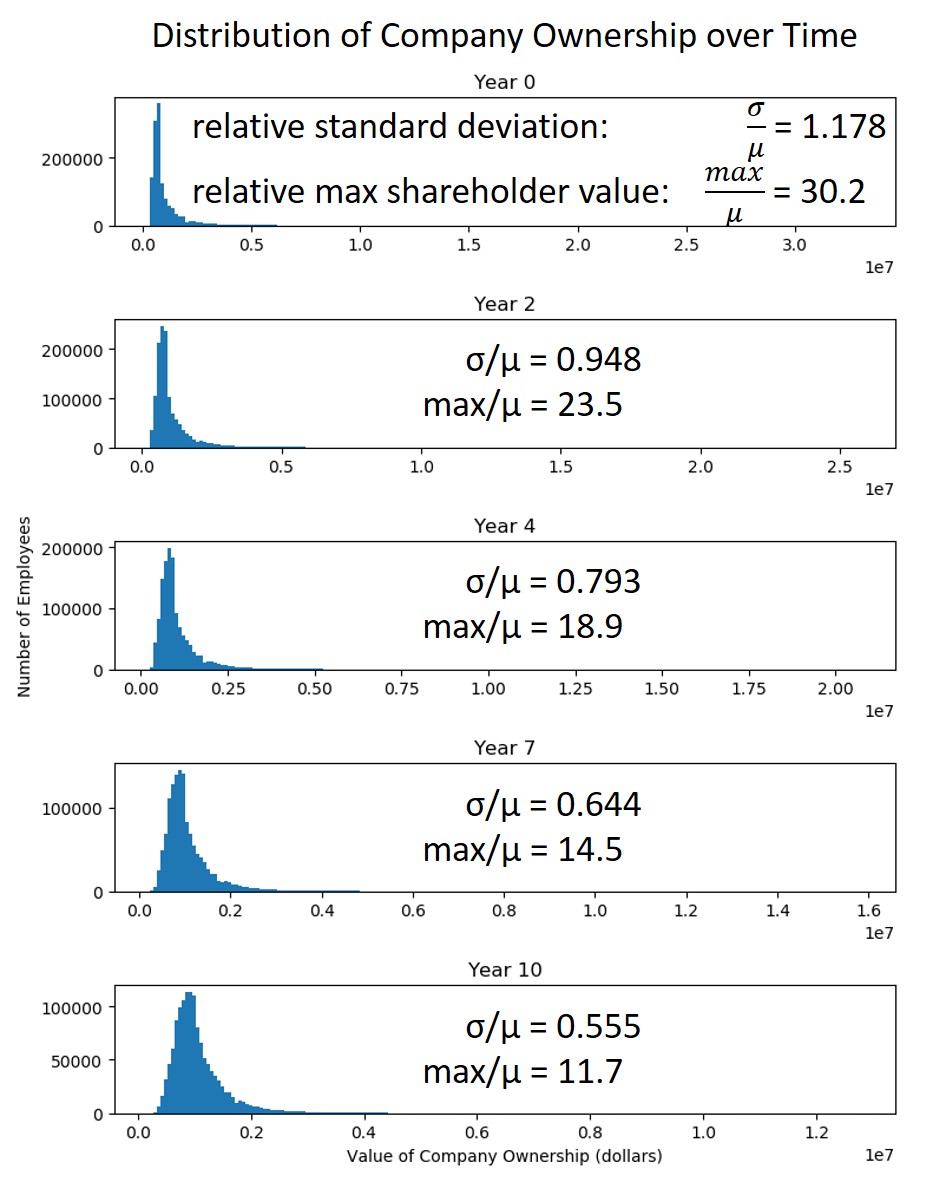}
    \caption{After 10 years, the vast majority of employees own a percentage of the company worth between \$50,000 and \$200,000.}
    \label{fig:img5}
\end{figure}

Using the parameters listed above, the relaxation of the system to its equilibrium state is simulated over the next 10 years.  The results of this simulation are shown in Figures \ref{fig:img4} and \ref{fig:img5}.  At the beginning of the simulation, the salary of the CEO was 30.2 times higher than that of the average  employee.  After 10 years, the relative salary of the CEO dropped from 30.2 to 3.7.

Another key metric, the relative standard deviation, is used to measure the compactness of the distribution.  A high relative standard deviation indicates the distribution is sparse, with many outliers far from the mean.  A low relative standard deviation means that the majority of the distribution can be found clustered in the neighborhood close to the mean.  During the simulation, the relative standard deviation decreased from 1.178 to 0.329, representing a more equitable salary distribution.

A similar trend occurs in the value of company ownership for each employee.  As most companies operate today, this value is zero by default.  Recoupling labor to ownership immediately increases the net worth of most employees.

At the beginning of the simulation, the ownership distribution is equal to the salary distribution.  As the salary distribution relaxes to equilibrium, the fraction of the company owned by each employee also becomes more equitable, as shown in Figure \ref{fig:img5}.  After 10 years, the majority of the 1,335,000 employees own a percentage of the company worth between \$50,000 and \$200,000.  The relative value of company ownership of the maximum shareholder, in this case the CEO, decreased from 30.2 times the value for the average employee to 11.7.  The relative standard deviation decreased from 1.178 to 0.555.

This chapter took the first step toward introducing the idea that, in order to accurately describe social processes, they must be viewed in terms of their energy and entropy content.  The first basic step taken to implement this new understanding, recoupling labor to ownership, immediately produced a major improvement in quality of life for the majority of individuals.

\subsection{Key Takeaways}
\begin{enumerate}
    \item Decoupling labor from ownership causes the modern capitalist system to be thermodynamically equivalent to an engine incentivized to output maximum energy at all times.
    \item This fact is the root cause of many of the world's problems.
\end{enumerate}

%% file: chaos01-Intro.tex
\twocolumn[
  \begin{@twocolumnfalse}
    \section{Chapter 2\\An Object-Relations Approach to Complex Systems}
    \vspace{2pt}
  \end{@twocolumnfalse}
]

\PARstart{H}{ow} can one observe the behavior of a system and determine its inner workings?  It is usually impossible to solve this problem exactly, since most real-world models contain uncertainty in relation to the systems they represent.  There is a tradeoff between model accuracy, model size, and computational cost.  There are often also hidden variables and elements of stochasticity.

Minimum Description Length (MDL) is often used as a parameter in model selection to fit a particular dataset, whereby the shortest description of the data, or the model with the best compression ratio, is assumed to be the best model \cite{rissanen1978modeling}.  MDL is a mathematical implementation of \textbf{Occam’s razor}, which dictates that among competing models, the model with the fewest assumptions is to be preferred.  If an MDL description can contain all of the operations in a Turing-complete programming language, the description length then becomes equivalent to the Kolmogorov complexity, which is the length of the shortest computer program that produces the dataset as output\cite{kolmogorov1965three}.  For a given dataset, a program is Pareto-optimal if there is no shorter program that produces a more accurate output.  The graph of Kolmogorov program length vs. accuracy is the Pareto frontier of a dataset.

\LinesNumbered
\RestyleAlgo{ruled}
\SetKwProg{Fn}{Function}{:}{end}
\SetKwInput{kwGet}{get}{}{}

If one combines Occam’s razor with the Epicurean \textbf{Principle of Multiple Explanations}, the result is Solomonoff's theory of inductive inference \cite{solomonoff1964formal1}\cite{solomonoff1964formal2}.   According to the Principle of Multiple Explanations, if more than one theory is consistent with the observations, all such theories should be kept.  Solomonoff's induction considers all computable theories that may have generated an observed dataset, while assigning a higher Bayesian posterior probability to shorter computable theories.  The theory of inductive inference uses a computational formalization of Bayesian statistics to consider multiple competing programs simultaneously in accordance with the Principle of Multiple Explanations, while prioritizing shorter programs in accordance with Occam's razor.

A characteristic feature of complex systems is their activity across a wide range of lengthscales and timescales.  As a consequence, they can often be divided into a hierarchy of sub-components.  In order to reduce the computational difficulty of programs that can be divided into sub-programs, the dynamic programming method, originally developed by Richard Bellman \cite{bellman1954theory}, simplifies a complicated problem by recursively breaking it into simpler sub-problems.   The \textbf{divide-and-conquer} technique used by dynamic programming can applied to both computer programming and mathematical optimization.  A problem is said to have optimal substructure if it can be solved optimally by recursively breaking into sub-problems and finding the optimal solution to each sub-problem.

In contrast to the divide-and-conquer procedure, a \textbf{unification} procedure finds a single underlying theory that can explain two or more separate higher-level theories.  Causal learning algorithms or structural learning algorithms are examples of unification procedures, which determine cause-and-effect among variables in an observed dataset.  These variables may be hidden or unknown, in which case they must be inferred via hidden Markov models, Bayesian inference models, or other methods.  

\begin{figure*}
    \centering
    \includegraphics[width=0.95\textwidth]{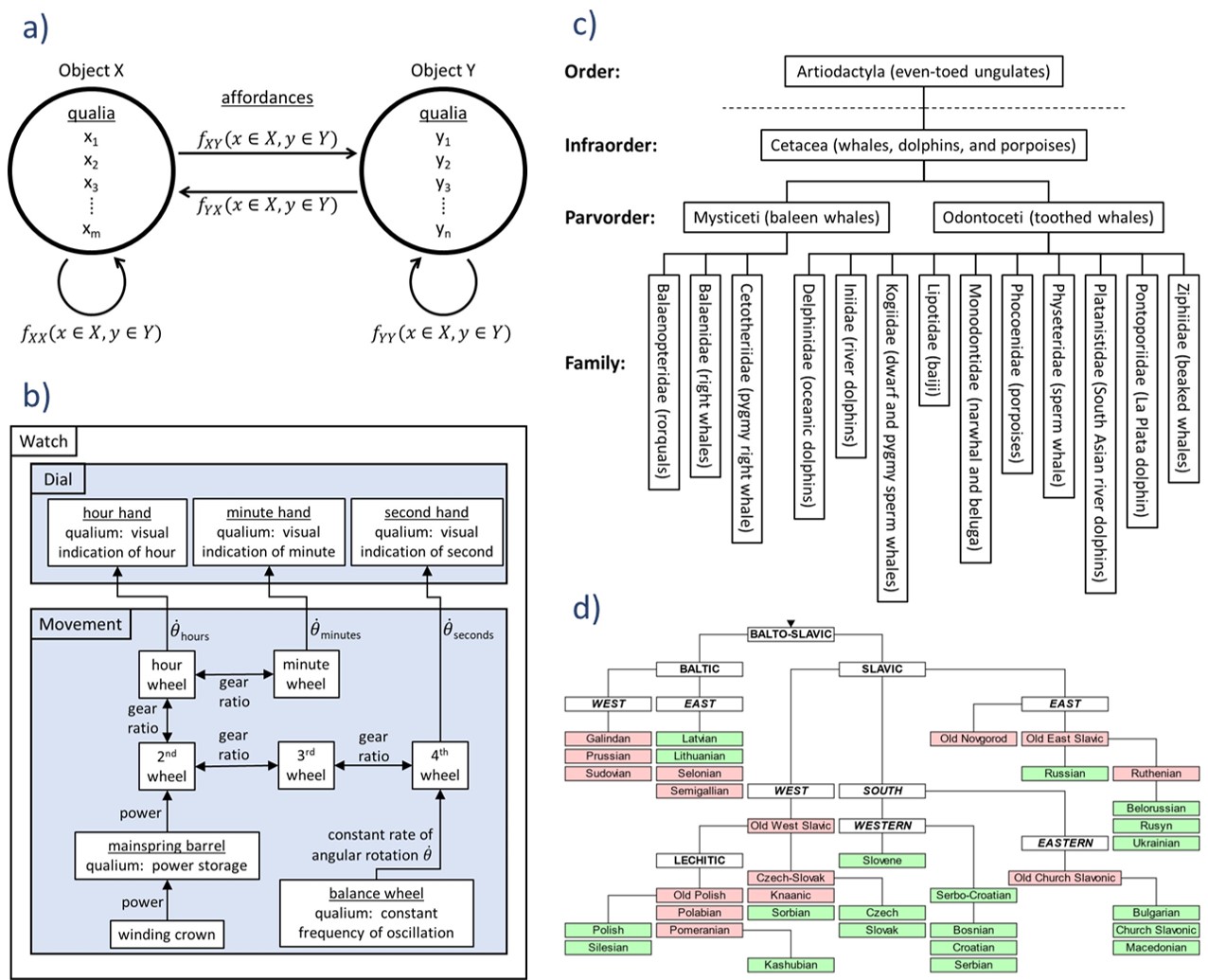}
    \caption{a) Objects contain qualia, which are properties, variables, or other objects.  Affordances are modes of interaction.  b) The relationships among objects can be shown using system diagrams.  In this example, a mechanical watch is composed of a dial object and a movement object.  The movement object is composed of gears, regulators, and springs, which affect each other through mechanical affordances.  c) The categories used for biological classification are not exact, but are hypothetical objects inferred by scientists and subject to revision.  This graph shows the categories for cetaceans between the ranks of order and family.   d) Connected objects in the graph share some amount of structure and mutual information.  In this language diagram, depicting the derivatives of the Balto-Slavic language which may have split into Baltic and Slavic around 1400 BCE\cite{gray2003language}, each language derives most of its information and structure from its predecessor in the graph.  Image licensed under CC BY-SA 3.0.}
    \label{fig:img58}
\end{figure*}

These four principles – Occam’s razor, the Principle of Multiple Explanations, divide-and-conquer, and unification – have been used in various combinations to analyze complex systems.  Some researchers have used them all together, such as Wu, Udrescu, and Tegmark in their endeavor to create an AI physicist\cite{wu2019toward}\cite{Udrescu_2019}\cite{udrescu2020ai2}.  In order to fully characterize the information contained in complex systems, a fifth item should be added to this list:  the principle of \textbf{overinterpretation}.  This principle states there can be multiple correct interpretations of the same dataset, and the validity of any particular lens of interpretation is determined by whether or not it produces useful insights.  This is distinct from the Principle of Multiple Explanations, which states that no plausible theory can be ruled out, but which still assumes there is one correct theory among many plausible.  In contrast to the Principle of Multiple Explanations, the principle of overinterpretation states that multiple interpretations can be simultaneously correct.  In the same way that the Principle of Multiple Explanations was suggested by Epicurus and refined by Bayes or Laplace, the principle of overinterpretation may be said to have been suggested by Freud in his analysis of Hamlet in The Interpretation of Dreams\footnote{“But just as all neurotic symptoms, and, for that matter, dreams, are capable of being ‘overinterpreted' and indeed need to be, if they are to be fully understood, so all genuinely creative writings are the product of more than a single motive and more than a single impulse in the poet's mind, and are open to more than a single interpretation.”\cite{freud1900interpretation}} and refined by Walter Kaufmann\cite{kaufmann1980}.  In a mathematical context, overinterpretation indicates there may be multiple non-equivalent Pareto-optimal objects within a given dataset.

\vspace{2.0ex plus .5ex minus .2ex}

%% file: chaos02-Theory.tex
\subsection{Theory}

The definitions used here are as general as possible.  The word \textbf{object} is defined as an entity containing two features:  1.  A set of properties, variables, or other objects, called \textbf{qualia} (the singular form used here is \textbf{qualium}), and 2.  functions between itself and other objects, called \textbf{affordances}.  A summary is shown in Figure \ref{fig:img58}a. 

An example for a mechanical watch is shown in Figure \ref{fig:img58}b.  The watch is composed of the dial and the movement, which are themselves composed of constituent objects.  Systems engineers and computer scientists will recognize the similarities to system design documents and Unified Modeling Language (UML) diagrams.

\subsubsection{Hierarchical Graphs}

In addition to representing objects using component diagrams or Venn diagrams, it is often useful to employ graph decompositions.  A taxonomy diagram for cetaceans is shown in Figure \ref{fig:img58}c, which separates general categories into more specific categories.  Similarly, Figure \ref{fig:img58}d shows the evolutionary diagram of the Balto-Slavic languages.  These graphs are hierarchical in the sense that each object derives much of its information and structure from its predecessor.

\subsubsection{Constructing and Deconstructing Objects}

What exactly constitutes an object?  Roughly, an object is an information-theoretic construct that contains a lot of information but does not require much description.  Shea\cite{Shea_2021} has applied this definition in physics by analyzing a fluid system containing both laminar and turbulent flow.  Without preprogrammed knowledge of different flow regimes, his machine learning algorithm discovered the boundary between the laminar and turbulent zones by minimizing the complexity of the model relative to its predictive capability.

\begin{figure}[ht!]
    \includegraphics[width=\linewidth]{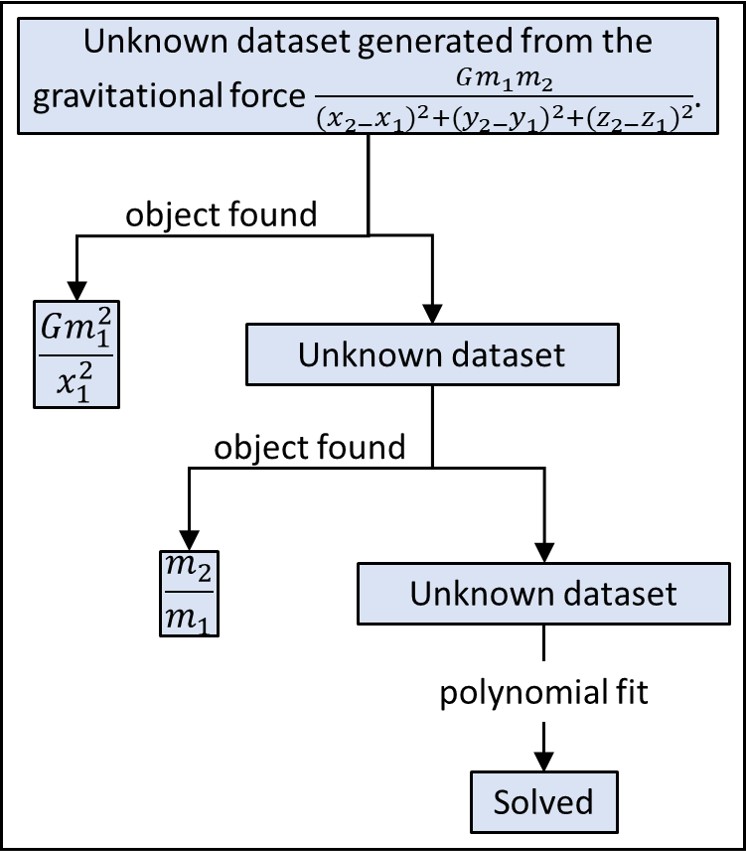}
    \caption{ML is used to procedurally find objects within a table of simulated gravitational measurements.  Image adapted from Figure 2 in \cite{Udrescu_2019}.  Used with permission.}
    \label{fig:img43}
\end{figure}

In simple algebraic systems, objects can be symbolic variables while affordances are mathematical functions.  Udrescu\cite{Udrescu_2019} has employed this formulation in the development of a machine learning algorithm to discover well-known physical laws from tables of experimental data.  Figure \ref{fig:img43} shows this process for Newton's law of universal gravitation $F = Gm_1m_2 / r^2$.  Given a table of $G,m,x,y$ and $z$ values, the program first discovers the dimensionless quantity $Gm_1^2/x_1^2$, then the ratio of masses $m_2/m_1$, then finally discovers the relationship among the position variables via polynomial regression.  The uncertainty present in the dataset is reduced each time a new object is found.

\paragraph{Structural Learning Methods}

Much previous work in learning the structural composition of systems has focused on Bayesian belief networks and hidden Markov models.  A Bayesian belief network is a probabilistic model that represents a set of variables and their conditional dependencies via a directed acyclic graph, meaning a graph in which information flows in only one direction along each edge, and which do not contain any circular causation loops.  These graphs are ideal for predicting the probable causes and contributing factors of observed events.  Algorithms exist for inferring unobserved variables, in addition to algorithms for belief propagation which can conditional probabilities throughout the graph in response to new evidence.

A hidden Markov model (HMM) assumes the system under investigation is a Markov process, which is a network of possible states of the system and the probabilities of transitions between states.  HMM's attempt to infer unobserved hidden states of the system.  The graphs used for Markov random fields can be undirected and contain cycles, which are more general than the directed acyclic graphs used for belief nets.  Consequently, HMM's are a useful analysis tool that have found applications across numerous fields.

While a great deal of progress has been made in structural inference algorithms such as Bayesian belief networks and hidden Markov models, they are limited in their ability to infer relationships among variables beyond conditional probabilities.  Conway's Game of Life is a good example.  The Game of Life is a well-known cellular automaton devised by John Conway\cite{gardner1970fantastic} which uses simple rules to produce complex and lifelike behavior on a two-dimensional grid of square cells using discrete timesteps.  The game applies the following rules each timestep:

\begin{enumerate}
    \item Any live cell with fewer than two live neighbors dies.
    \item Any live cell with two or three live neighbors lives.
    \item Any live cell with more than three live neighbors dies.
    \item Any dead cell with exactly three live neighbors becomes a live cell.
\end{enumerate}

Although these rules are simple, they will defeat a probabilistic inference algorithm's attempt to discover them.  This is because the behavior of a cell is not stochastic in relationship to its neighbors, but instead is based on the application of basic logical operations.  In order to solve this system, something more akin to Wu, Udrescu, and Tegmark's AI physicist is needed.

\paragraph{The Role of Machine Learning}
\label{section:ml}

\begin{figure}[t!]
    \includegraphics[width=\linewidth]{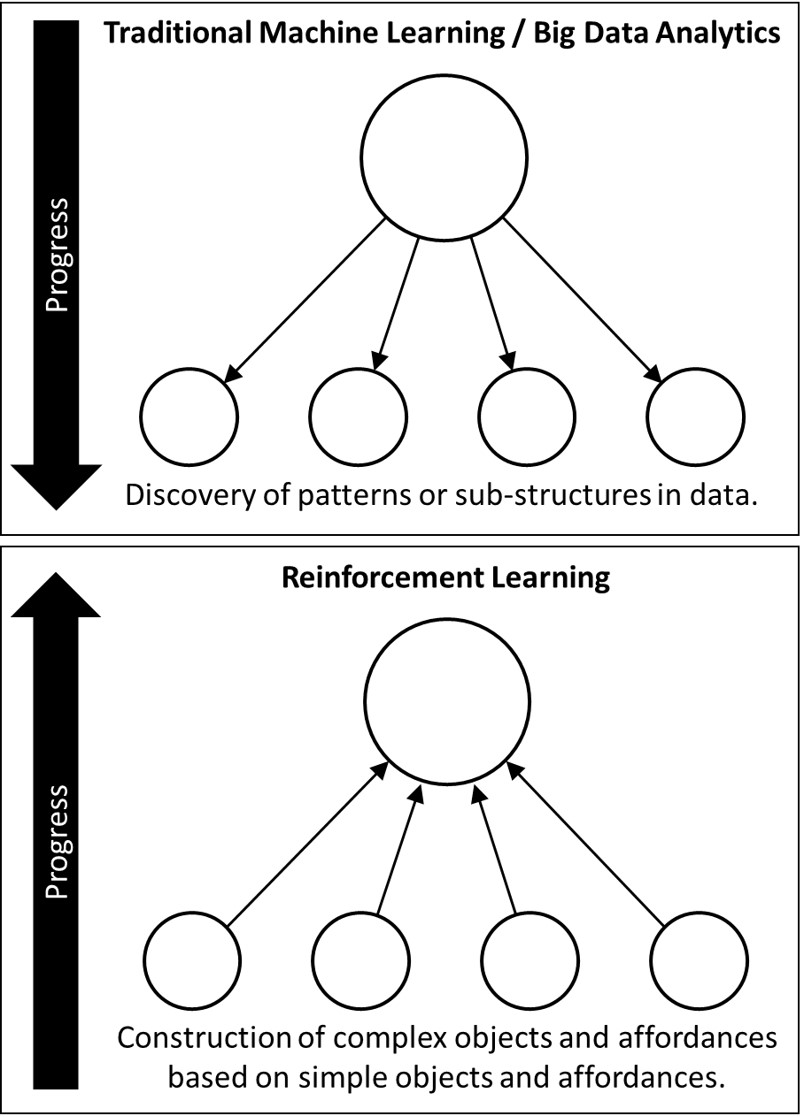}
        \caption{Machine learning involves either the deduction of objects present in datasets (top) or the construction of complex objects from simple constituents (bottom).}
    \label{fig:img44}
\end{figure}

Continuing with the Game of Life example, a more robust machine learning (ML) algorithm can be devised to infer objects and affordances at across a range of length- and timescales.  The various subtypes of ML can be broadly divided into two categories.  Casual references to machine learning or data science typically refer to the first category, termed here as traditional ML or big-data analytics.  This category involves analyzing large datasets using some combination of statistical methods to uncover previously unknown patterns or relationships among variables.  Image recognition would be included in this group, as the neural network must analyze pixels in an image to determine unknown variables, such as whether the image contains a dog or a cat.

\begin{algorithm}[t]
\caption{Zoom-in-Zoom-out (ZIZO)}\label{alg:zizo}
    \SetKwFunction{Fzo}{Zoom\_out}
    \SetKwProg{Fn}{Function}{:}{\KwRet $X$}
    \Fn{\Fzo{Y}}{
        \tcc{This function takes a top-down approach from the macroscale.}
        View the system as a black box, and infer the hidden objects X that could be causing its behavior\;
        If Y is provided, condition this process on Y\;
    }
    \SetKwFunction{FDeduce}{Zoom\_in}
    \SetKwProg{Fn}{Function}{:}{\KwRet $Y$}
    \Fn{\FDeduce{X}}{
        \tcc{This function takes a bottom-up approach from the microscale.}
        Encode the microstructure of the subsystem into representative objects Y\;
        If X is provided, condition this process on X\;
    }
	\SetKwFunction{FMain}{}
	\SetKwProg{Pn}{Main}{:}{\KwRet}
	\Pn{\FMain{}}{
	    \While{$X \neq Y$}{
            X = Zoom\_out(Y)\;
            Y = Zoom\_in(X)\;
    	    Loop until convergence\;
        }
    }
\end{algorithm}

The second subtype of machine learning is reinforcement learning (RL), wherein an agent interacts with its environment in order to achieve a goal or set of goals.  Much RL development has focused on games, such as chess, go, and classic Atari video games.  In these environments, the agents use a finite set of actions, such as chess moves or video game inputs, to progress toward achieving a long-term goal.  In this process, RL agents often discover advantageous techniques or game configurations already known to human experts, then proceed to surpass human experts by discovering previously unknown techniques.  In contrast to the traditional ML category, the RL category involves the construction of advanced objects and affordances from simpler objects and affordances.  This dichotomy is illustrated in Figure \ref{fig:img44}.

\begin{figure}[b!]
    \includegraphics[width=\linewidth]{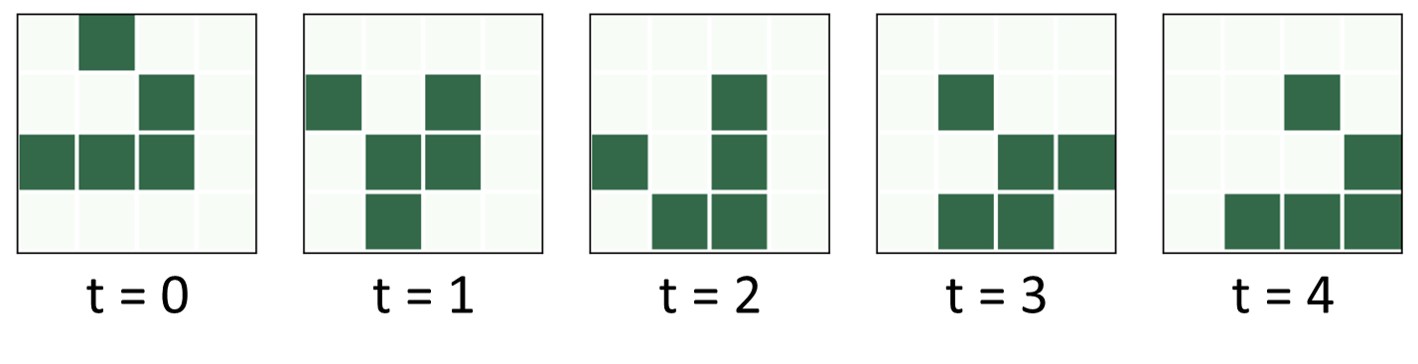}
    \caption{The glider in Conway's Game of Life is a pattern that moves diagonally down and right.  When viewed from the macroscale, the glider is an object moving across the game board.  However, when viewed on the microscale, it's a collection of tiles turning on and off.  Image licensed under Creative Commons\cite{glider}.}
    \label{fig:img55}
\end{figure}

The glider in Figure \ref{fig:img55} is one of the of the simple forms found in the Game of Life.  As illustrated in the figure, the glider moves one space down and to the right every five timesteps.  The glider appears to be moving across the game board when viewed from a distance.  However, when viewed at the level of individual cells, motion is not possible as the cells can only turn on and off.  The cells do not have the affordance of motion, but this affordance appears through the collective action of multiple objects.  This emergence of complex affordances from combinations of simple affordances is the process denoted by the bottom half of Figure \ref{fig:img44}.  As with many physical systems, there is a difference in the behavior and understanding of the glider at the microscale compared to the macroscale.

Direct parallels can be drawn between the simple Game of Life example and problems in physics and engineering.  For example, the prediction of macroscopic chemical properties based on molecular structure is challenging due to the complex behavior caused by quantum effects on the microscale.  A similar problem exists in neuroscience in the attempt to bridge the gaps between the behavior of individual neurons and the brain en masse.

In order to address this problem, we have defined an algorithm to discover the connections among objects across different lengthscales and timescales.  Inspired by the Grow-Shrink algorithm used for Bayesian network structure learning and the forward-backward algorithm used for Markov models, here we have defined a Zoom-in-Zoom-out (ZIZO) algorithm for object-relations models.  The Grow-Shrink algorithm works by adding variables to nodes in a Bayesian network graph until all dependencies have been captured (the ``grow" phase), then tests these dependencies and removes superfluous connections (the ``shrink" phase)\cite{margaritis2003learning}.  The forward-backward algorithm for HMM's calculates the probability distribution of the system's hidden state variables for a sequence of observations moving forward in time, then repeats the process for the sequence in reverse, with the intention to cause the forward-in-time model and the backward-in-time model to converge\cite{rabiner1986introduction}.  The ZIZO algorithm performs a similar type of oscillation by first observing a phenomenon from a zoomed-out, macroscopic viewpoint and developing a top-down model, then zooming into an object on the microscopic level and building a bottom-up model, then iterating between the two models until they converge on a common set of objects shared between the two.

Note the ZIZO algorithm has so far only been discussed as a method to describe complex systems.  However, if the algorithm is used in a reinforcement learning mode, and the program tries to create desired affordances rather than trying to model observations, then the algorithm changes from a tool of system description to a tool of system design.  This type of tool could be used to autonomously design and optimize all types of engineering systems, perhaps beyond the capabilities of human experts.  The introduction of intelligent sampling via active learning, in addition to simulation capabilities, could further improve the utility of this process.

\subsubsection{A Simple Example:  A Connected Graph with 8 Nodes}

\begin{figure*}
    \centering
    \includegraphics[width=1\textwidth]{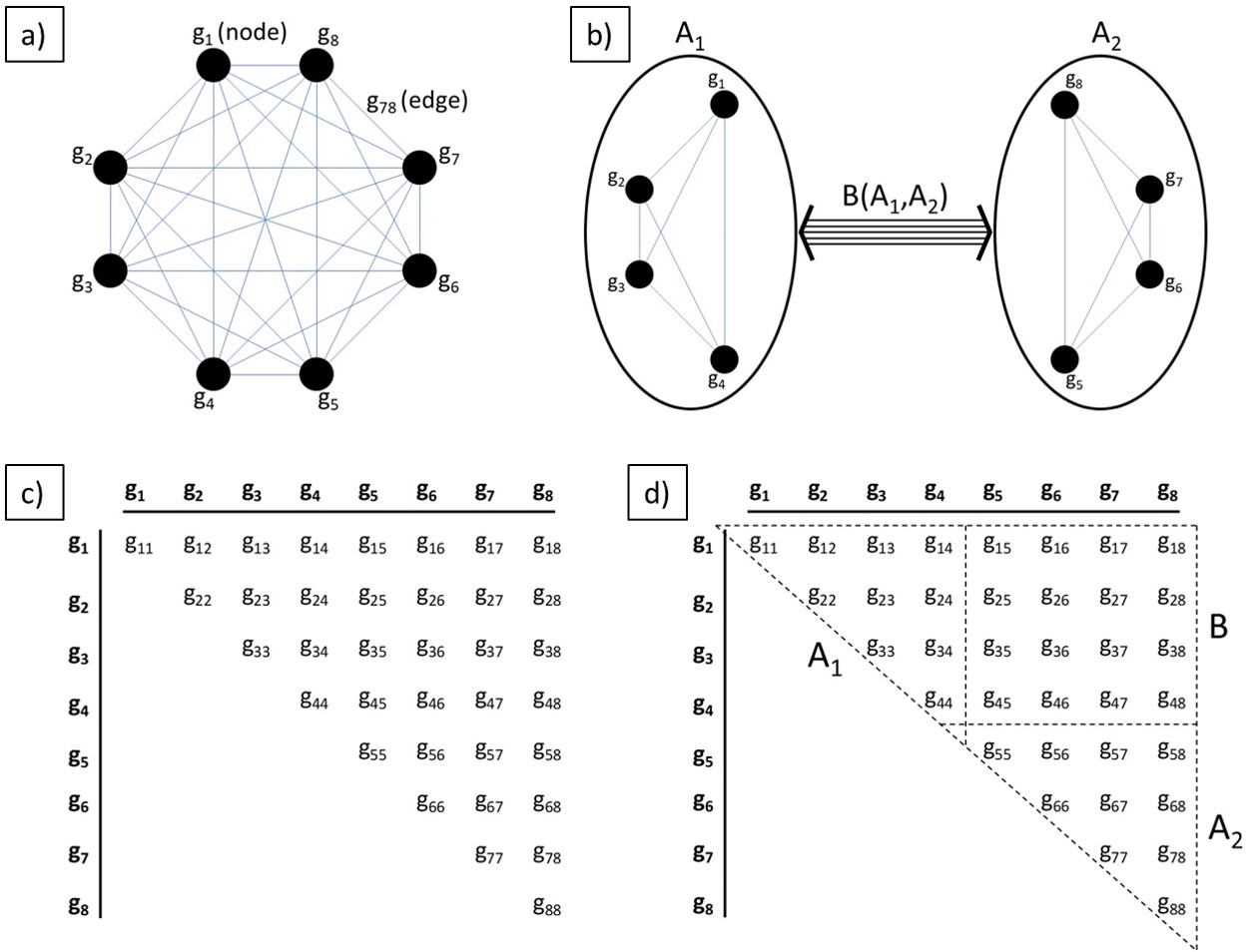}
    \caption{a) Graph $G$ is composed of nodes $g_1$ through $g_8$ and possible edges $g_{ij}$ between nodes $g_i$ and $g_j$ for $i,j \in [1,8]$.  b) Dividing $G$ into two objects $A_1$ and $A_2$ produces two k-connected graphs of size 4 and a function $B$ that communicates between them.  c) As $g_{ij}=g_{ji}$ the adjacency matrix of $G$ is symmetric about the diagonal.  d) Dividing $G$ into $A_1$ and $A_2$ has the effect of segmenting the adjacency matrix.}
    \label{fig:img62}
\end{figure*}

As an intuitive introduction to the theory of delineating objects, we will define a graph $G$ as a k-connected graph of size 8.  There are 8 nodes, each of which may or may not share an edge with any of its neighbors or with itself.  We will assume edges are bidirectional, meaning edge $g_{12}$ is the same as edge $g_{21}$.  Figure \ref{fig:img62}a illustrates this graph.

There may or may not be an edge $g_{ij}$ between nodes $g_{i}$ and $g_{j}$.  If there is an edge, this condition is represented by $g_{ij}=1$.  If not, $g_{ij}=0$.  We will assume no prior knowledge about the existence of each edge, meaning $g_{ij}$ may be 0 or 1 with equal probability.  The full information contained in $G$ is therefore represented by the adjacency matrix shown in Figure \ref{fig:img62}c.

The adjacency matrix contains 36 independent binary variables.  This translates to $2^{36}$ equally probable configurations of $G$.  The Shannon entropy of $G$ is therefore 
\begin{equation}
    H(G) = \mathrm{log}_2 2^{36} = 36 \mathrm{\ bits}.
\end{equation}
Intuitively, information entropy can be considered the average amount of information conveyed by an event when considering all possible outcomes.

Let us now consider what happens if $G$ is not regarded as a single object, but is instead analyzed as two objects $A_1$ and $A_2$ as shown in Figure \ref{fig:img62}c.  $A_1$ is constructed to contain nodes 1-4, while $A_2$ contains nodes 5-8.  An affordance $B$ describes the possible modes of interaction between $A_1$ and $A_2$.

Using the boundaries drawn in this example, $A_1$ and $A_2$ are k-connected graphs of size 4.  Since $A_1$ contains nodes 1-4, the self-entropy of $A_1$ is the entropy contained in the edges $g_{ij}$ for $i,j \in [1,4]$.  Likewise, the self-entropy of $A_2$ is the entropy contained in the edges $g_{ij}$ for $i,j \in [5,8]$.  The function $B$ must contain information about the connections between nodes 1-4 and nodes 5-8, therefore $B$ contains the values of $g_{ij}$ for $i \in [1,4]$ and $j \in [5,8]$.
\\

\begin{figure*}
    \centering
    \includegraphics[width=1\textwidth]{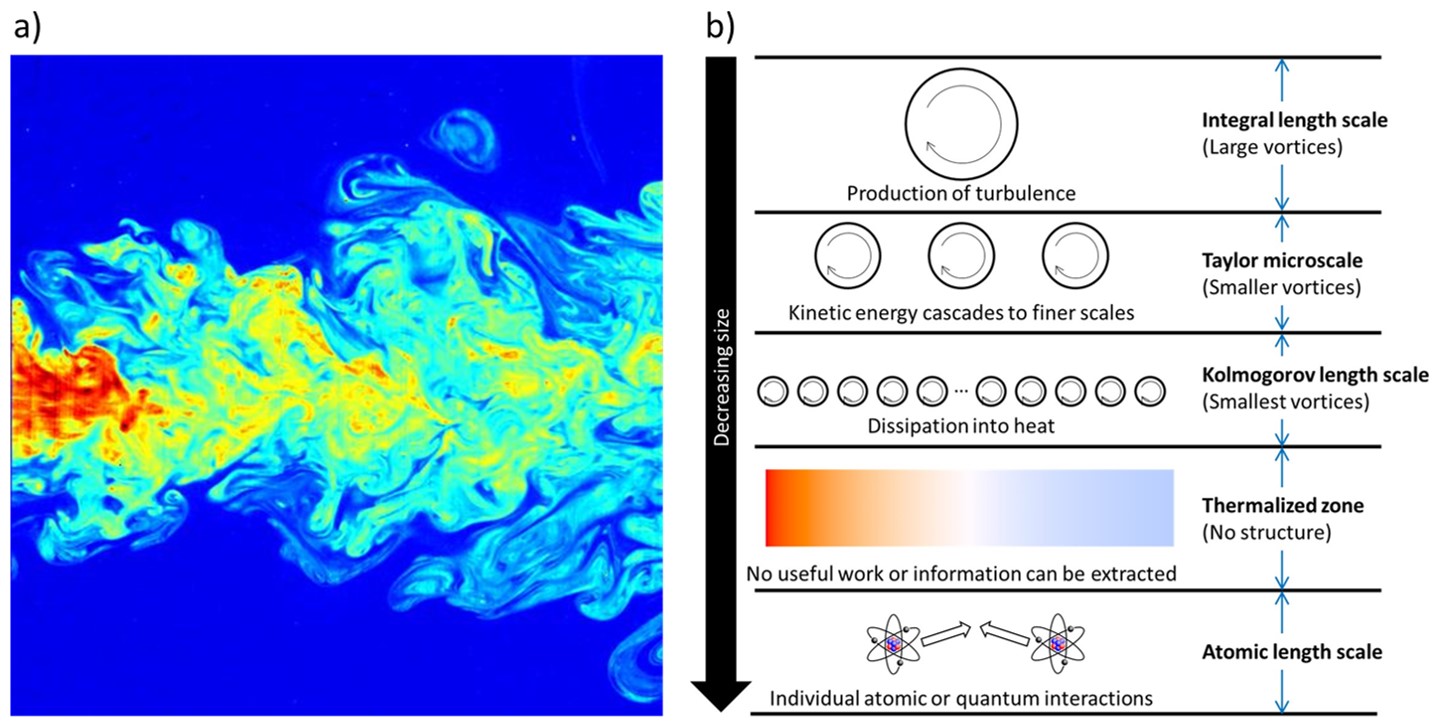}
    \caption{a) False-color flow visualization of a submerged turbulent jet, made by laser-induced fluorescence.  Imaged licensed under CC BY-SA 3.0\cite{turb}.  b) The CIE of turbulence is high due to presence of structures across man lengthscales and timescales.}
    \label{fig:img60}
\end{figure*}

These groups are visualized on the adjacency matrix in Figure \ref{fig:img62}b.  The separation of $G$ into $A_1$ and $A_2$ had the effect of dividing the adjacency matrix into sections.  The information content of each section can be analyzed separately:
\begin{align}
    H(A_1) = \mathrm{log}_22^{10} = 10\\
    H(A_2) = \mathrm{log}_22^{10} = 10\\
    H(B) = \mathrm{log}_22^{16} = 16.
\end{align}
Total information in the system is conserved:
\begin{equation}
    \label{eqn:eqn1}
    H_{\mathrm{total}} = H(G) = H(A_1)+H(A_2)+H(B) = 36.
\end{equation}
Conditional entropy equalities also hold.  For example,
\begin{equation}
    H(A_1) = H(G) - H(G \mid A_2,B).
\end{equation}

\paragraph{Discussion}
In this example, it's clear dividing $G$ into two connected objects is equivalent to considering $G$ as a single object.  This can be verified by the equality in Equation \ref{eqn:eqn1} or by visually inspecting Figure \ref{fig:img62}d.  In fact, the boundaries around $A_1$ and $A_2$ can be drawn around any set of nodes and the outcome will be equivalent in regard to the total system entropy.  The only change will be how the adjacency matrix is divided into subsections.

From an intuitive standpoint, the separation of $G$ into $A_1$ and $A_2$ seems undesirable, as a large amount of information needs to be contained in the connection $B$.  $B$ contains 16 bits of information, while the objects $A_1$ and $A_2$ each contain only 10.  Separating $A_1$ and $A_2$ does not decrease the amount of information needed to describe the system.

However, what if the values in $B$ are almost always zero?  In this case, the entropy contained in $B$ is very low.  If nodes 1-4 are almost never connected to nodes 5-8, it becomes advantageous to consider graphs $A_1$ and $A_2$ as separate groups.

\subsubsection{Complex Information Entropy}

Here we will define a useful method to measure the information entropy of any system.  The Complex Information Entropy (CIE) of a system is defined as

\begin{equation}
    \label{eqn:cie}
    \begin{gathered}
        \textrm{Complex Information Entropy } \mathnormal{}C^*(S) = \\
        \sum_{A_i \in \mathbb{A^*}}c_iH(A_i) + \underset{B_{ij} \in \mathbb{B}^*\mathnormal{}}{\sum_{A_i, A_j \in \mathbb{A^*}, \ \mathnormal{}}}d_{ij}H(B_{ij}(A_i, A_j))
    \end{gathered}
\end{equation}

\noindent where $\mathbb{A}^*$ and $\mathbb{B}^*$ are the sets of all Pareto-optimal objects and affordances in system $S$, $A_i$ and $B_{ij}$ are defined in Equation \ref{eqn:po}, and $c_i$ and $d_{ij}$ are empirically-determined coefficients used to adjust for information not captured by $A_i$ or $B_{ij}$.  This function is recursive, as each object $A_i$ may itself be considered a complex system, and $H(A_i)$ can be calculated as the CIE of the objects and affordances inside $A_i$.  Thus the CIE function accounts for the hierarchical representation of information in complex systems.

$C^*$ indicates the optimum value of $C$ given a complete set of optimized objects and affordances.  Similar to the uncomputability of Kolmogorov complexity, meaning there is no finite-length program that will return the Kolmogorov complexity for a given input string, $C^*$ may also be uncomputable.  In practice, real-world programs may be used to iteratively find better values for $C$ over time.

Pareto-optimal objects and affordances are those that optimally simplify some subset $S'$ of $S$, as given by the equation

\begin{equation}
    \label{eqn:po}
    \begin{gathered}
        \mathnormal{}C(S' \subseteq S) = \\
        \underset{A_i \in \mathbb{A}, \ \mathnormal{}B_{ij} \in \mathbb{B}}{\mathrm{argmin}} \sum_{i}c_iH(A_i) + \sum_{i,j}d_{ij}H(B_{ij}(A_i, A_j))
    \end{gathered}
\end{equation}

\noindent where $\mathbb{A}$ and $\mathbb{B}$ are the sets of all possible objects and affordances in system $S$, including those that are sub-optimal.

\subsubsection{A More Complex Example:  Turbulence}

Equation \ref{eqn:cie} will now be applied to turbulent flow.  The best way to model, understand, and quantify the information contained in turbulence is an open question in physics.  The CIE metric provides useful insights when applied to such systems.

Turbulence possesses the property of structure at multiple scales.  Instabilities in the fluid flow form vorticies, which create secondary instabilities and smaller vorticies, which create even smaller vorticies, in a process that cascades to smaller length scales until the kinetic energy dissipates into heat.  The formation of these eddies can be seen in Figure \ref{fig:img60}a.

Researchers investigating the behavior of turbulent flow have identified three distinct physical regimes:  the integral length scale, the Taylor microscale, and the Kolmogorov length scale, as shown in Figure \ref{fig:img60}b.  The integral length scale is the regime in which eddies form and gather energy.  Below the integral length scale is the Taylor microscale in which energy is transferred from larger vortices to smaller vortices without being lost to the surrounding fluid.  Smallest is the Kolmogorov lengthscale, in which the viscosity of the fluid causes the energy of the vortices to be dissipated as heat.  There is effectively no structure below this scale, as the fluid has thermalized into a maximum-entropy condition characterized by the Maxwell-Boltzmann distribution from which no information can be extracted.

In addition to the energy differences, the three regimes also differ in their isotropic characteristics.  The integral length scale is highly anisotropic, meaning the eddies favor certain orientations in 3-dimensional space.  The Kolmogorov scale is isotropic, meaning the eddies occur evenly in every direction.  The transition from anisotropy to isotropy occurs in the Taylor microscale.

A lot of information is needed to fully describe a turbulent system due to the differences in each regime.  The CIE value for a system using the integral-Taylor-Kolmogorov schema is

\begin{equation}
    \label{eqn:eqn2}
    \begin{gathered}
        C(\mathrm{turbulence}) = \\ 
        C(\mathrm{Kolmogorov}) + C(\mathrm{Taylor}) + C(\mathrm{integral }) = \\
        \sum_{A_i \in \mathrm{Kol.}}c_iH(A_i) + \sum_{A_i, A_j \in \mathrm{Kol.}}d_{ij}H(B_{ij}(A_i, A_j)) + \\
        \sum_{A_i \in \mathrm{Kol.}, A_j \in \mathrm{Tay.}}d_{ij}H(B_{ij}(A_i, A_j)) + \\
        \sum_{A_i \in \mathrm{Tay.}}c_iH(A_i) + \sum_{A_i, A_j \in \mathrm{Tay.}}d_{ij}H(B_{ij}(A_i, A_j)) + \\
        \sum_{A_i \in \mathrm{Tay.}, A_j \in \mathrm{int.}}d_{ij}H(B_{ij}(A_i, A_j)) + \\
        \sum_{A_i \in \mathrm{int.}}c_iH(A_i) + \sum_{A_i, A_j \in \mathrm{int.}}d_{ij}H(B_{ij}(A_i, A_j)).
    \end{gathered}
\end{equation}

Equation \ref{eqn:eqn2} may appear to be unwieldy,  however its modularity is a major strength.  The CIE of each layer may be calculated separately.  Regardless of the number of layers in the hierarchy, each layer only needs to consider the interactions between itself and the layers immediately above and below.  Moreover, each object only needs to consider itself and the interactions with its immediate neighbors.

\begin{figure}[t]
    \includegraphics[width=\linewidth]{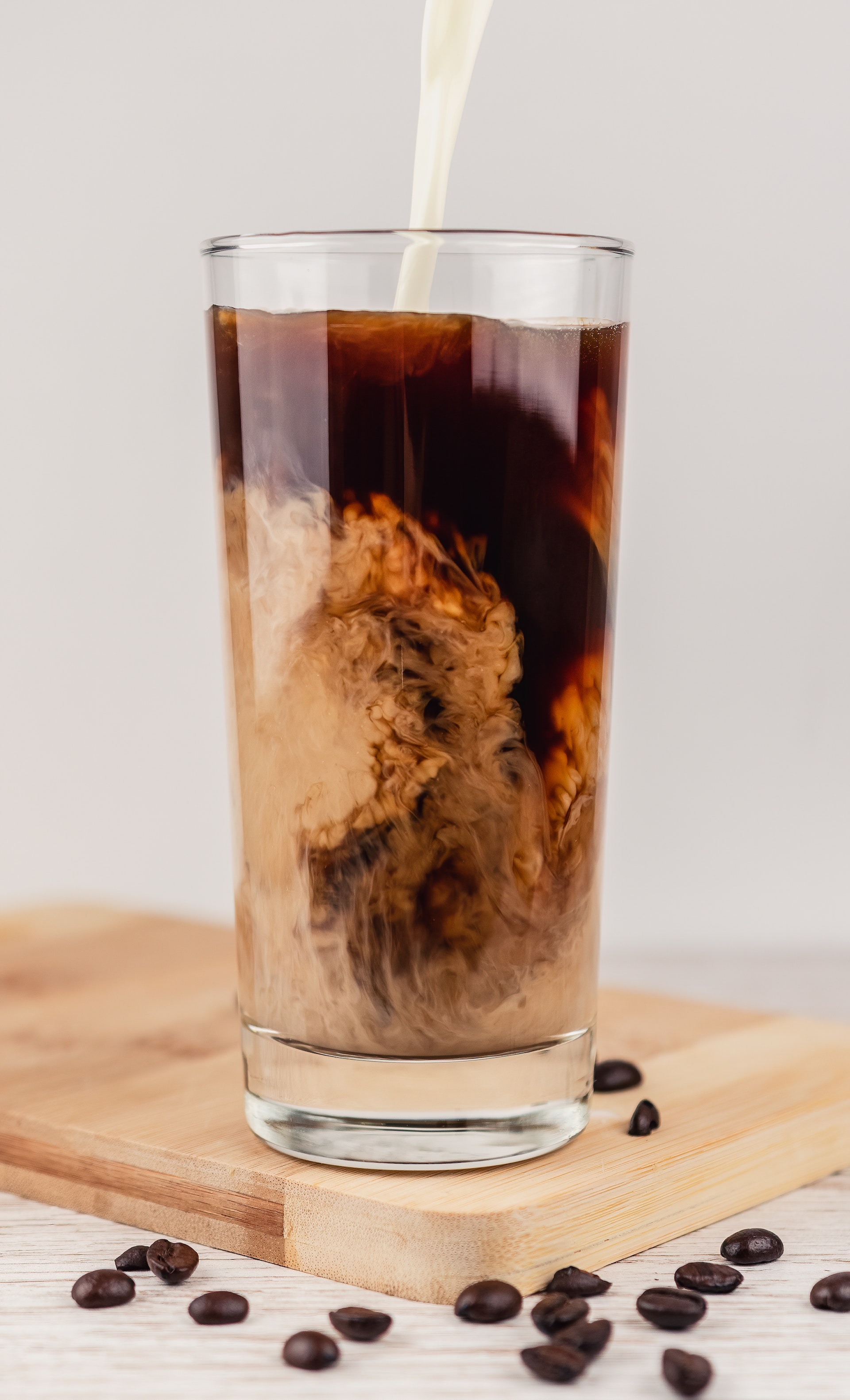}
    \caption{Mixing cream into coffee creates turbulent eddies at multiple lengthscales.  The complexity of the system is greatest when the two are partially mixed, although the classical Gibbs entropy is maximized when the mixture is homogeneous and without structure.  Imaged licensed under CC\cite{coffee}.}
    \label{fig:img41}
\end{figure}

As a sanity check, CIE behaves as desired for the classical example of cream being mixed into black coffee.  The Gibbs entropy used in statistical mechanics is maximized when the liquids are fully mixed and contain no internal structures, while the complexity is maximized during the mixing process when turbulent structures are present as shown in Figure \ref{fig:img41}.  As expected, CIE is low during the unmixed and fully-mixed states, and high during the partially-mixed state.

\subsubsection{The Virus Effect:  Propagation of Entropy through a System}

\begin{algorithm*}[hbt!]
\caption{Complex System Structure-Learning Algorithm}\label{alg:structure_learning}
    \SetKwFunction{FMain}{}
	\SetKwProg{Pn}{Main}{:}{\KwRet}
	\Pn{\FMain{}}{
        \While{not converged}{
    	    Search for objects and affordances in the system\tcc*[r]{\textbf{divide-and-conquer}}
    	    Verify found objects and affordances are Pareto-optimal\tcc*[r]{\textbf{Occam's razor}}
    	    Add all Pareto-optimal objects and affordances found to a list:\tcc*[r]{\textbf{overinterpretation}}
    	    Check if objects in the list can be merged by using a common substructure\tcc*[r]{\textbf{unification}}
    	    Consider each object on the list to be its own system and repeat the above steps\tcc*[r]{\textbf{recursion}}
    	    Graph building:  search for dependencies among objects\;
    	    \If{If new dependencies are found}{
                Modify the objects and affordances if needed\;
    	    }
    	    Graph pruning:  check to see if existing dependencies can be removed:\;
    	    Loop until convergence, or repeat indefinitely\;
    	}
	}

\end{algorithm*}

One feature of complex systems is the ability for small-scale changes to propagate throughout the system and cause large-scale effects.  For example, a person's exposure to a single virus, which is much smaller than a cell, can lead to a cascade of effects that overwhelm their entire body.  How can this be analyzed in an object-relations model?

The answer is found in the state entropy of each object.  Using the example of a virus, one may imagine the human body as consisting of a hierarchical arrangement of organs, tissues, and cells.  The physical processes and relevant length- and timescales for small components, such as cells, are much different than those for large components, such as organs.  Accordingly, the operation of organs is not affected by individual cells, and it is not necessary to model individual cells in order to accurately model organs.

If the output and states of object $A$ are discrete, the entropy of $A$ is given by the equation

\begin{equation}
    \label{discrete_entropy}
    H(A) = -\sum_{a \in \mathscr{A}} p(a)\mathrm{ln}\mathnormal{}(p(a)) = E[-\mathrm{ln}\mathnormal{}(p(A))]
\end{equation}

\noindent where $a$ is the internal condition of $A$, $\mathscr{A}$ is the set of all possible values of $a$, and $p(a)$ is the probability of occurrence of $a$.  If the states of $A$ are continuous, the entropy is given by

\begin{equation}
    \label{continuous_entropy}
    H(A) = E[-\mathrm{ln}\mathnormal{}(f(A))] = -\int_{\mathscr{A}} f(a)\mathrm{ln}\mathnormal{}(f(a))da.
\end{equation}

\noindent Here $\mathscr{A}$ is the support of $A$, meaning the domain over which $a$ is nonzero, and $f(A)$ is the probability density function of $A$.

If the scheme for representing the information in object $A$ is efficient, then the states that occur most frequently will carry the lowest information entropy, while the most unusual states will carry the highest information entropy.  In general, if $p_s$ is the probability of object $A$ being in state $s$, the entropy of object $A$ in state $s$ is $H(A(s)) = -p_s \mathrm{ln} (p_s)$.  In a living system such as a cell, the states expected to occur most frequently are ideally the states in which the cell is functioning as usual without any major problems.  If the cell is attacked by a deadly virus, the cell will be placed in a highly unusual state, which will carry a high amount of information $H(A)$.  If cell $A_i$ infects cell $A_j$ via $B_{ij}(A_i,A_j)$, then the information entropy of $A_j$ will also increase.

In this optimal multi-scale model of the body, if a hypothetical infection remains local, then the high-CIE state will also remain limited.  However, if the high-CIE state propagates to larger lengthscales, then tissues, organs, and eventually the entire body will be affected.  If the full body is in an abnormally high-CIE state, this could signal a medical emergency.

This type of evaluation could hypothetically be used to identify one harmful virus out of millions of harmless viruses.  Notably, such a model remains computationally tractable, as each object only needs to consider the interactions with the objects to which it is directly connected.

\subsubsection{Implications for Biological Organisms}
\label{section:biology}

\begin{figure}[b]
    \includegraphics[width=\linewidth]{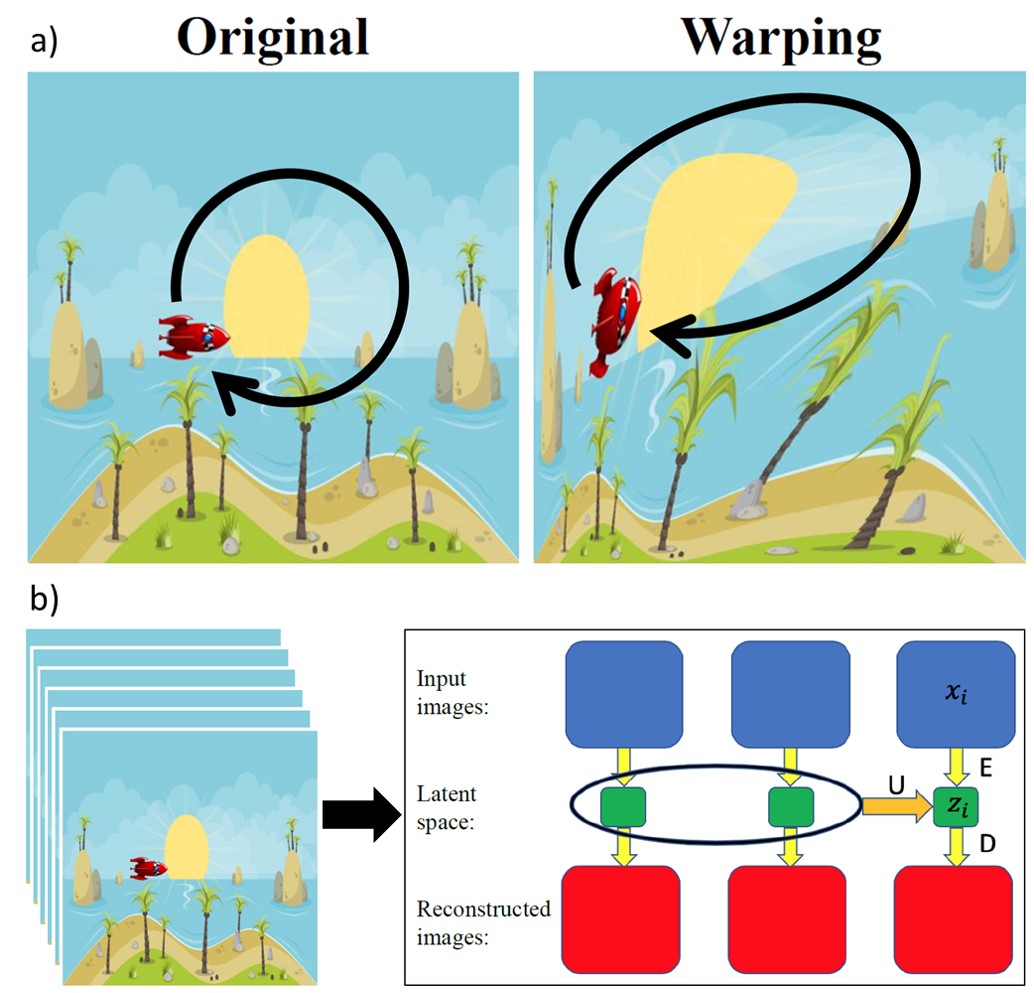}
    \caption{Using an autoencoder on distorted video resulted in the recovery of the original equations of motion for the rocket.  It is conjectured that humans perform a similar type of autoencoding of our perceptions in order to simplify our model of the world while retaining important information.  Image adapted from Figures 1, 2, and 3 in \cite{Udrescu_2021}.  Used with permission.}
    \label{fig:img50}
\end{figure}

\begin{figure*}
    \centering
    \includegraphics[width=.9\textwidth]{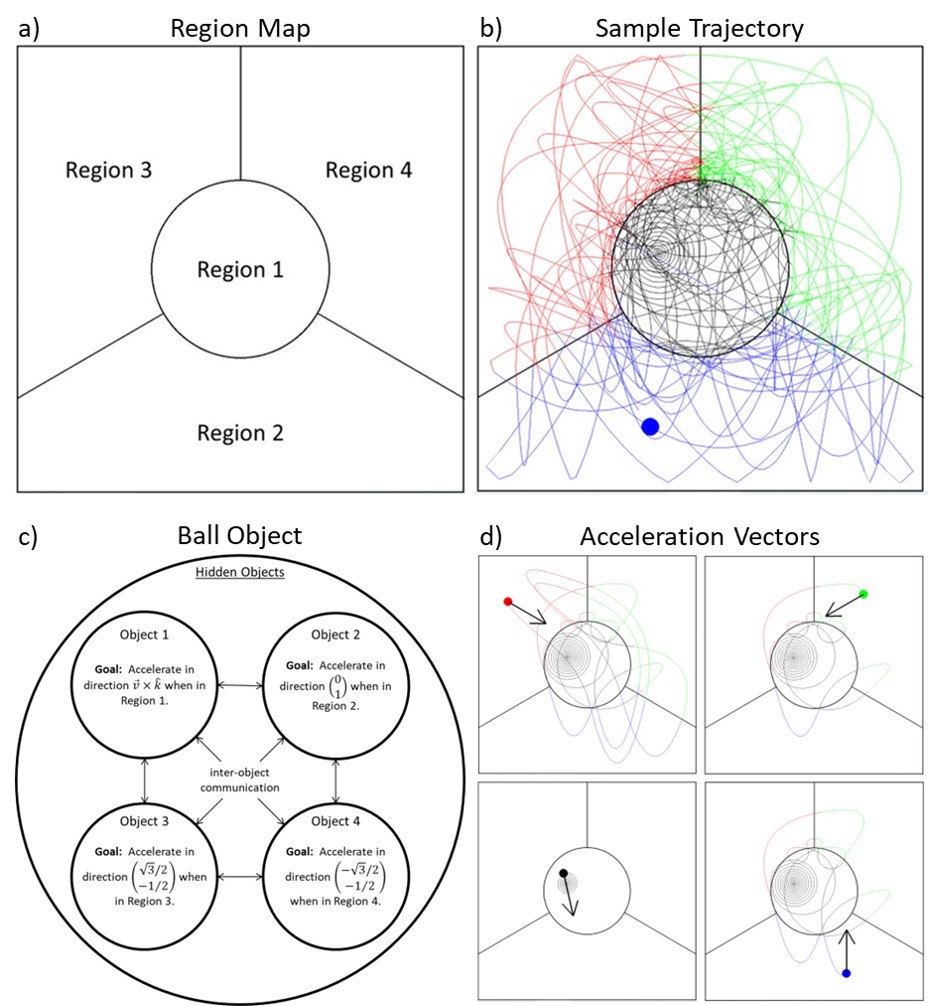}
    \caption{a) The map is divided into four regions.  The ball behaves differently in each region depending on which object is active.  b) The trajectory path is assigned a different color in each map region.  c) Four hidden objects work simultaneously to determine the ball's movement.  d) Acceleration vectors in each of the four regions.  The vector directions are constant in the edge regions and variable in the center.}
    \label{fig:img61}
\end{figure*}

The nervous systems of biological organisms seem to perform something conceptually similar to Equations \ref{eqn:cie} and \ref{eqn:po}, by creating a model of the world that is maximally descriptive while simultaneously being as simple as possible.  This is the Good Regulator theorem of cybernetics, which states that every good regulator of a system - meaning a device that maintains itself in a desired state - must contain a model of that system\cite{conant1970every}.  Sensory inputs are processed by the nervous system in order to construct an internal model of the world and to inform the organism's actions.  This is equivalent to the $\underset{A_i \in \mathbb{A}, \ \mathnormal{}B_{ij} \in \mathbb{B}}{\mathrm{argmin}}$ function of Equation \ref{eqn:po}, which finds the ideal choice of objects to represent a particular system.  The survival of an organism depends on its ability to recognize important information, disregard unimportant information, and update its model of the world in response to new observations.

Udrescu \cite{Udrescu_2021} has performed a simulation related to this idea.  He started by developing a video of a flying rocketship with a beach and palm trees in the background as shown in Figure \ref{fig:img50}.  The rocket moves around the frame by following trajectories determined by the equations for physical forces such as gravity, electromagnetism, and harmonic oscillation.  The video was then warped and distorted to obscure the rocket's motion.  Udrescu was able to recover the rocket's equations of motion by feeding the video frames into an autoencoder E, applying an operator U to the autoencoder's latent space, and training the operator and autoencoder simultaneously to achieve the simplest possible equation for U.  As a result, the autoencoder effectively learned how to unwarp the image, while the operator learned the symbolic equations of motion in a standard Cartesian coordinate system.

The implication of this experiment is that the human nervous system performs something similar to the function of the CIE equation:  developing a model of the world that is maximally descriptive while simultaneously being maximally efficient.  The following section will implement a practical example of such a model using trajectory data.


%% file: chaos03-Methods.tex

\subsection{Methods}

\begin{algorithm}[hbt!]
\caption{Deduction of Sub-Objects}\label{alg:deduction}
    \tcc{This program will try to continue finding objects in the dataset until its model of the data stops improving.}

    \SetKwFunction{FDeduce}{separate\_objects}
    \Fn{\FDeduce{dataset}}{
        Train a neural network on the complete dataset as a baseline\;
    	Randomly split the dataset into two subsets\;
    	Train two new NN's on each subset\;
    	\While{NN losses are decreasing}{
    	    Test each data point using both new NN's\;
    	    Reassign each data point to the set of the NN that fits best\;
    	    Retrain both NN's\;
    	    Loop until convergence\;
    	    \tcc{This is an annealing process.  Data points will migrate to subsets that minimize the combined losses from NN1 and NN2.}
    	}
    	\eIf{new NN losses $<$ baseline NN loss}{
    	    \KwRet the new subsets, new NN's, and total NN losses\;
    	    \tcc{New objects were found.}
    	}{\KwRet the original dataset, baseline NN, and baseline NN loss\;
    	    \tcc{New objects not found.}
    	    }
	}

	\SetKwFunction{FMain}{}
	\SetKwProg{Pn}{Main}{:}{\KwRet}
	\Pn{\FMain{}}{
	    \textbf{get} dataset\;
        \While{total NN losses are decreasing}{
            \ForEach{subset}{
    	        \textbf{call} separate\_objects(\textit{subset})\;
    	        \If{new objects are found}{
    	            Update the subsets for each object\;
    	        }
	        }
    	    Loop until convergence\;
        }
    }
\end{algorithm}

The theory described above will now be used to discover hidden laws of motion within a set of trajectory data.  A general method for learning the structure of complex systems is shown in Algorithm \ref{alg:structure_learning}. This problem may be formulated as a reinforcement learning task and/or an optimization task, where the goal is to find the Pareto-optimal object boundaries.  Starting with some object $A$ with boundary $a = \partial A$, the goal is to iteratively adjust $a$ in order to Pareto-optimize all objects $A$ and affordances $B$.  The Bellman value equation for this problem takes the form

\begin{equation}
       C(s) = \underset{a = \partial A}{\mathrm{min}} \underset{s = \{A,B\}}{\sum} R(s,a,s') + \gamma C(s')
\end{equation}

\noindent where the state $s$ is the current state of the system model, $s'$ is the state from the previous iteration, $R$ is the reward for changing the model from state $s'$ to $s$ by changing the boundary $a$, $C(s')$ is the Complex Information Entropy from the previous state, and $\gamma$ is a factor which accelerates convergence\cite{bellman1957}.  Because the function is minimizing rather than maximizing $C$, $\gamma$ is greater than 1.  The corresponding update rule is
\begin{equation}
    C_{i+1}(s) \leftarrow \underset{a = \partial A}{\mathrm{min}} \underset{s = \{A,B\}}{\sum} R(s,a,s') + \gamma C_i(s').
\end{equation}
\noindent Ideally, this will produce an optimal policy for adjusting $a$.  In the following example, an annealing procedure is used as the method to adjust $A$ into an optimal configuration.

\begin{figure*}
    \centering
    \includegraphics[width=1\textwidth]{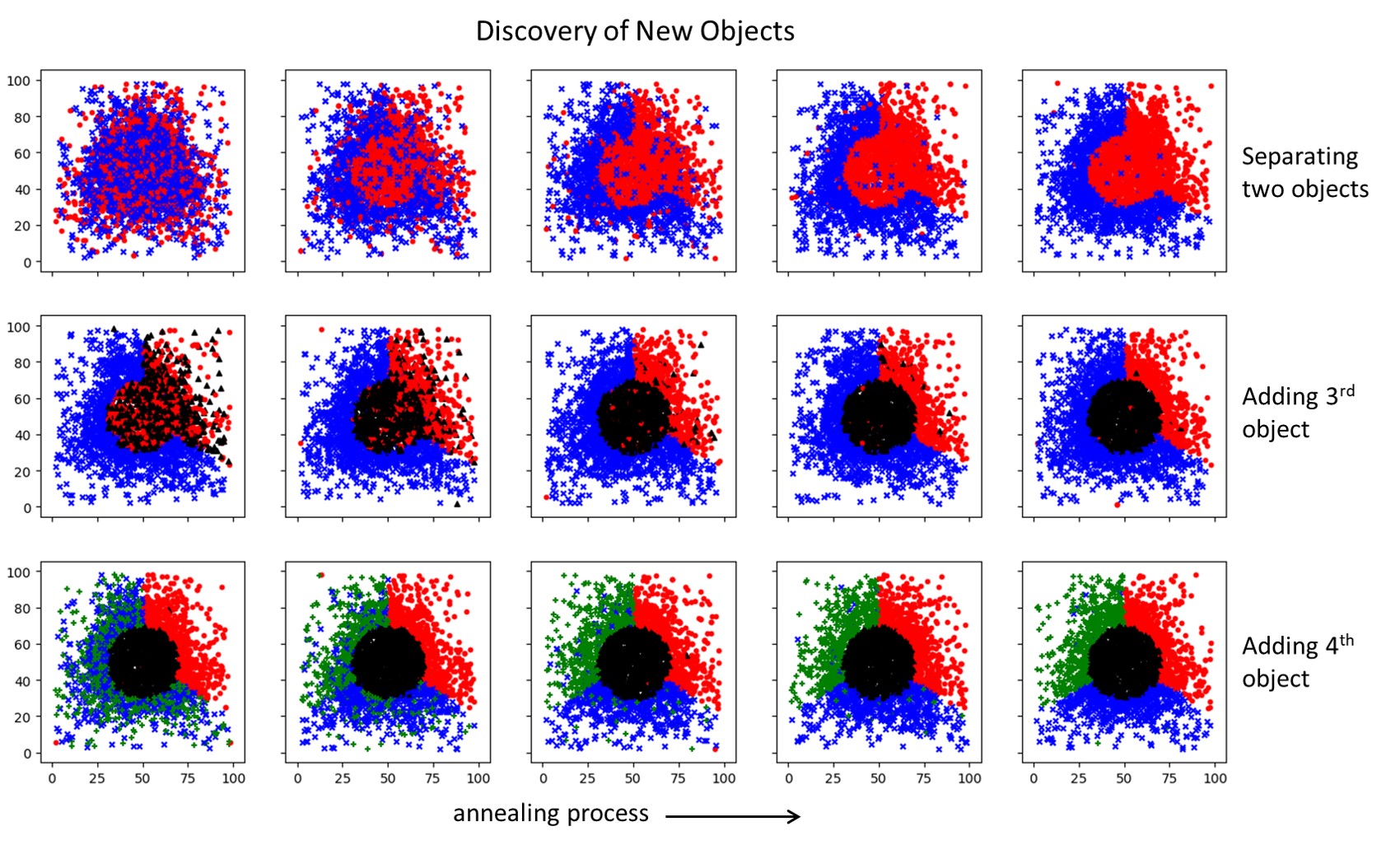}
    \caption{The machine learning algorithm is able to infer the existence of all four objects, along with their regions of operation.}
    \label{fig:img30}
\end{figure*}

\subsubsection{Inferring Hidden Objects through Observation}

We begin with a square map divided into four regions as shown in Figure \ref{fig:img61}a.  A ball is placed at a random location on the map with a velocity vector pointed in a random direction with magnitude close to zero.  The ball then moves in a particular trajectory, depending on its region.  If the ball is in region 1, it accelerates in direction $\vec{v} \times \hat{k}$.  The direction of the acceleration vector in Region 2 is $\begin{pmatrix} 0\\ 1 \end{pmatrix}$, in Region 3 is $\begin{pmatrix} \sqrt{3}/2 \\ -1/2 \end{pmatrix}$, and in region 4 is $\begin{pmatrix} -\sqrt{3}/2 \\ -1/2 \end{pmatrix}$.  The magnitude of the acceleration is constant.  A sample trajectory is shown in Figure \ref{fig:img61}b.

In this simulation, external forces do not act on the ball.  Instead, the ball has a constant internal supply of energy and moves of its own volition depending on the region in which it finds itself.  The ball contains four hidden objects, each with a set of movement instructions, as shown in Figure \ref{fig:img61}c.  Communication between the hidden objects is used to allocate the ball's fixed supply of energy.

After running the simulation to generate the ball's trajectory, the $x$ and $y$ coordinates of the ball at each timestep can be used as a dataset to train a machine learning algorithm. Algorithm \ref{alg:deduction} shows the method used to infer the presence of hidden objects.  Each time a hypothetical new object is added, the performs an annealing process refine the model of each object.  The program continues to add objects until the addition of a new object does not improve the model's accuracy.

%% file: chaos04-Results.tex

\subsection{Results}

\begin{figure}[b]
    \includegraphics[width=\linewidth]{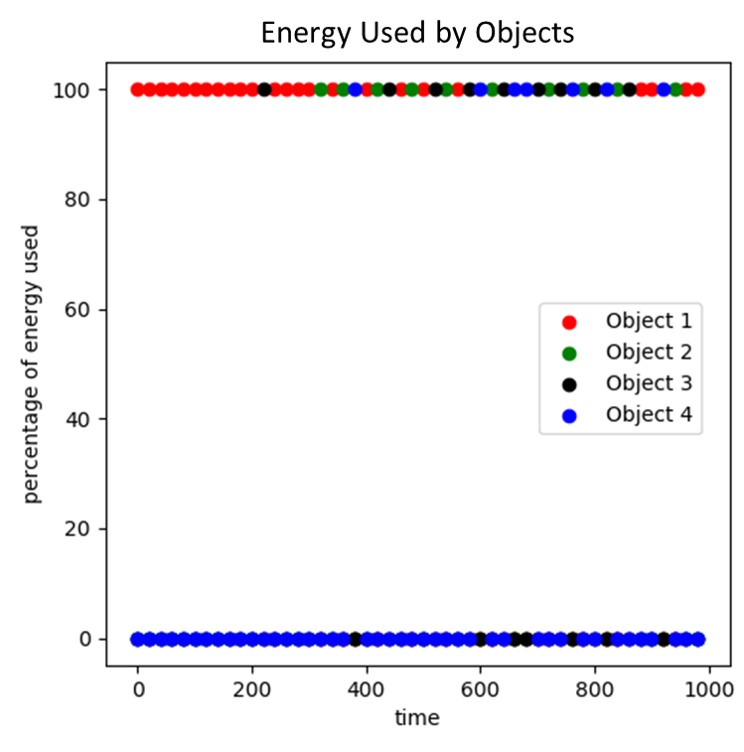}
    \caption{The object activations can be used to verify and gain insight into the model.  Only one object is active at a time during the course of the simulation.}
    \label{fig:img56}
\end{figure}

The annealing process is visualized in Figure \ref{fig:img30}.  As shown in the figure, the program first discovers two suboptimal objects.  The program learns the correct spatial boundaries while determining the region of applicability for each object.  Notably, the regions of the datasets correlated to each object are connected spatially, but not temporally.  The ball's trajectory enters and exits each region multiple times during the simulation.  The algorithm does not enforce set connectedness or convexity during the annealing process, as seen in the starting positions on the left side of Figure \ref{fig:img30}.  The datapoints belonging to each set are initially dispersed randomly throughout the map.  They are iteratively sorted into their correct positions until the algorithm's error reaches a local minimum.

After this minimum is reached, the program hypothesizes the existence of a third object in the region of maximum error, which is the region indicated by red circles in the top row of Figure \ref{fig:img30}.  Predictably, the region containing the center of the map contains the maximum error, as the ball's motion in the center is more complex than near the edges.  The program repeats the annealing process on the second rows of Figure \ref{fig:img30} and correctly identifies the objects corresponding to Regions 1 and 4 in Figure \ref{fig:img61}a.

In the final iteration, shown on the third row of Figure \ref{fig:img30}, the program correctly identifies the objects corresponding to Regions 2 and 3.  The error in this state is the global minimum error of the system.  Adding a fifth object will not improve the model, nor will an attempt to unify existing objects as described on line 6 of Algorithm \ref{alg:structure_learning}.

We can also check the efficiency of the solution by observing the activation of the individual objects.  If we assume the ball has a fixed supply of energy to be distributed to the various objects, an accurate solution will show only one object active at each moment.  Figure \ref{fig:img56} shows the allocation of energy during the simulation.  As expected, the energy used by each object is either 100\% or 0\%, indicating that only one object is active at a time. The force vector in each region is shown in Figure \ref{fig:img61}d.

\subsection{Discussion}

\subsubsection{Environmental Effects on an Object}

\begin{figure}[t]
    \includegraphics[width=\linewidth]{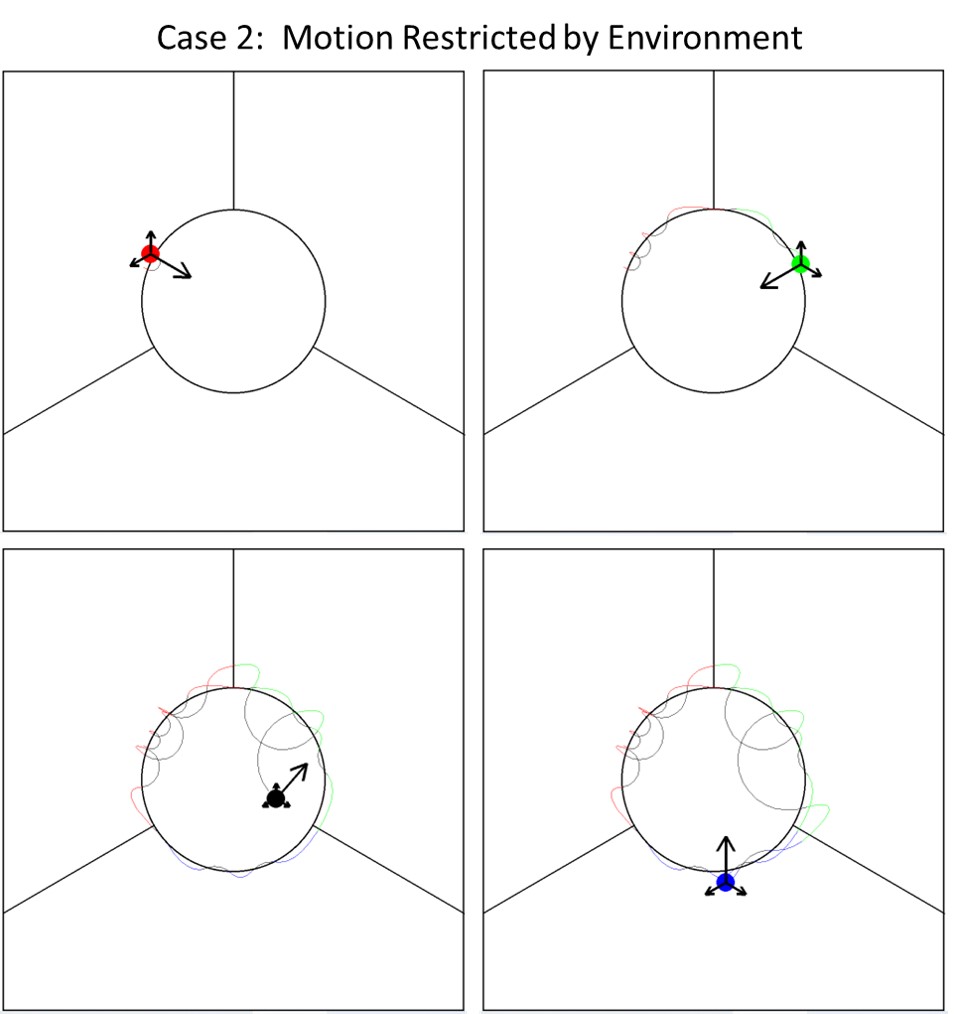}
    \caption{The acceleration vectors when the distance travelled each timestep is limited.  Objects act against each other in order to comply with the restriction imposed by the environment.}
    \label{fig:img33}
\end{figure}

\begin{figure}[b]
    \includegraphics[width=\linewidth]{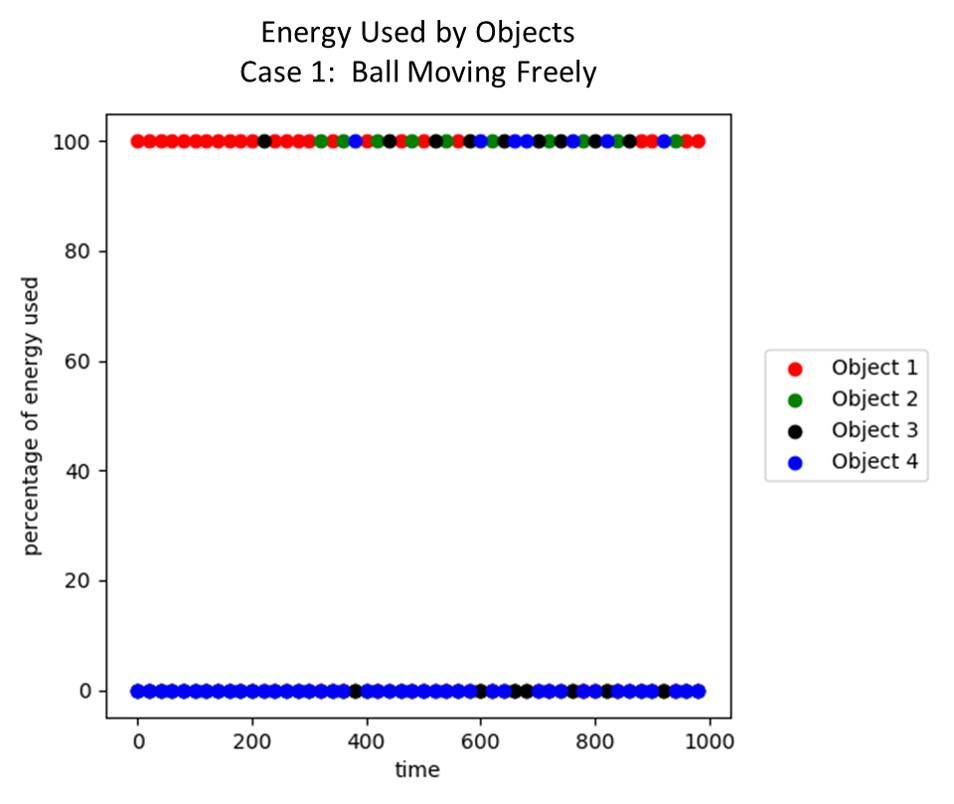}
    \caption{Energy allocation under ideal conditions.  100\% of energy is used for useful work.}
    \label{fig:img34}
\end{figure}

We will now add a complication to the simulation.  In the first stage of the simulation, the ball is allowed to move freely without restriction from the environment.  In the second stage, an external restriction is imposed upon the ball.  If the ball wants to move a distance $D$ during timestep $\Delta t$, the environment will not allow the ball to move a distance greater than $D/2$.  The environment is not applying an opposing force, but is rather imposing a condition that a move by a distance greater than $D/2$ is not possible or not allowed.  As in the first stage of the simulation, the only energy source is the internal energy of the ball.

The unrestricted acceleration vectors in each of the four regions are shown in Figure \ref{fig:img61}d.  A fascinating effect happens when the environmental restriction is applied:  the other objects, which are normally inactive, wake up and begin to influence the ball's motion.  They effectively fight against each other.  Instead of one object being active at a time, multiple objects are recruited simultaneously, each supplying its own motion vector.  In regions 2, 3, and 4, the primary object accelerates forward while the other two objects collaborate to slow it down.  In region 1, the primary object receives 50\% of the available energy, while the other three objects each receive 16.7\%, causing their motion vectors to sum to zero.  These scenarios are shown Figure \ref{fig:img33}. 

\begin{figure}[b]
    \includegraphics[width=\linewidth]{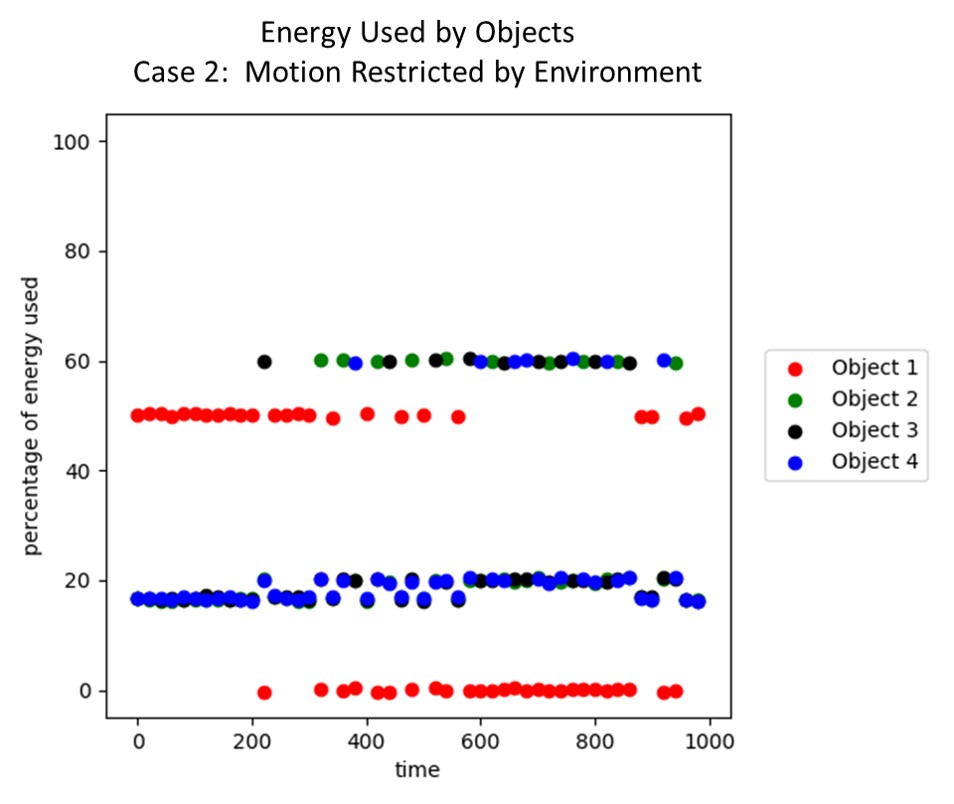}
    \caption{Energy allocation under restricted conditions.  50\% of available energy is used for useful work.  The other 50\% is wasted as heat or some other form of inefficiency.}
    \label{fig:img35}
\end{figure}

As shown in Figure \ref{fig:img61}c, the four objects are modelled as nodes in a message-passing network, which can pass information to each other about their energy usage.  Because there is a fixed amount of energy, the objects must allocate energy amongst themselves in order to adapt to the restriction imposed by the environment.  The result of this energy allocation is shown in Figures \ref{fig:img34} and \ref{fig:img35}.  In the free-moving scenario, the energy used by each object is either 100\% or 0\%.  None of this energy is wasted; it is all converted into motion.  In the restricted scenario, the energy utilization of each object is between 0 and 60\%.  Only 50\% of this energy is converted into motion.  In a real-world system, the other 50\% would be converted either into heat or into other methods of energy expenditure without useful work.

\paragraph{Conjecture 1:  Self-Interference}

\begin{conj1}
    \label{con:self-destruct}
    When not functioning optimally, complex systems begin causing interference to their own operation.
\end{conj1}

\paragraph{A Resolution of the Dark Room Paradox}

As discussed in Section \ref{section:biology}, a biological organism naturally develops a descriptive and efficient internal representation of its environment.  The dark room paradox is a problem that arises in cognitive science when one assumes the organism is incentivized to construct an internal model of its world that is as accurate as possible.  Maximizing the accuracy of a predictive model is equivalent to minimizing the surprise that will be experienced when an event occurs, which is equivalent to minimizing the information entropy of observation.  Continuing with this line of reasoning, an organism could bring its predictive error to zero by sitting in a dark, silent room and not observing anything.

The paradox is removed if one considers an organism's highest goal is to maximize CIE rather than minimize prediction error.  For a given system, equation \ref{eqn:cie} automatically minimizes prediction error by specifying the optimum objects $A$ and affordances $B$.  Maximizing $C$ is equivalent to the organism developing the richest and most complete understanding possible of its world, which includes minimizing prediction error as a subcomponent.

\paragraph{Objects in the Human Psyche}

\begin{figure}[b]
    \includegraphics[width=\linewidth]{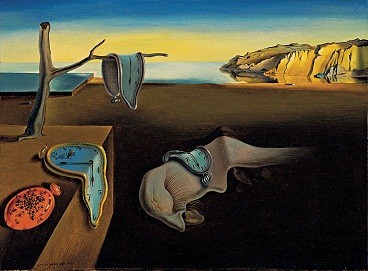}
    \caption{The Persistence of Memory by Salvador Dalí.}
    \label{fig:img36}
\end{figure}


When an individual produces a work of art, their creation contains, or implies, much more information than is needed to encode the pixels in an image or the letters in a text.  A book, image, song, video, or other media may be compressed into a trivially small file on a computer, containing very little Shannon entropy, but still may be tremendously meaningful.  How can this be?

\begin{figure*}
    \centering
    \includegraphics[width=\linewidth]{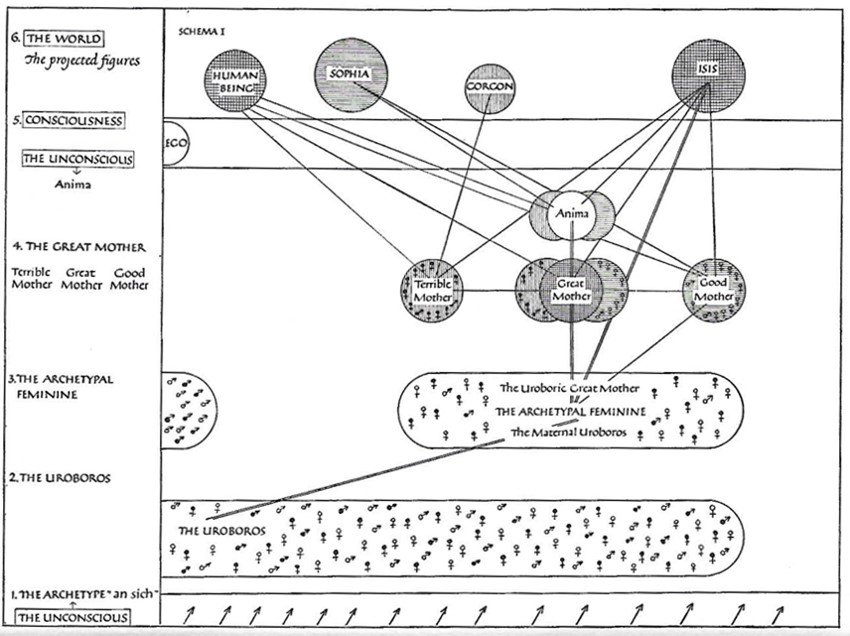}
    \caption{The progression of development of objects in the psyche deriving from the mother object.  From ``The Great Mother:  An Analysis of the Archetype" by Erich Neumann \cite{neumann2015great}.  Used with permission.}
    \label{fig:img38}
\end{figure*}

The answer has to do with the internal objects in the psyche of the viewer.  When an individual produces a work of art, they are, in some fundamental sense, acting as a thermodynamic information refinement engine.  The information produced in art or literature is of a sophisticated quality, since an individual viewing the artwork can infer a great deal about the creator and his environment.  When an individual views a work of art, the interaction between the viewer and the media may be understood as the effects of the media on the viewer's internal psychic objects.  An object in this context is any structure in the psyche that meets the criteria for $A$ and $B$ in equation \ref{eqn:cie}.  This very general definition my include, but is not limited to, internal representations of people, places, things, ideas, archetypes, complexes, or other structures.  Therefore, although the entropy of the bits needed to store the text of a novel may be low, the CIE of the novel in the psyche of the reader may be exceptionally high.

Interestingly, in the imagination and dreams, qualia and affordances may be mixed and recombined among objects in ways that are not possible in the physical world.  The Persistence of Memory \ref{fig:img36} is a good example.  In this image, the watches possess the qualium of viscosity, which is not possible physically.

The ability of the psyche to reallocate qualia and affordances is very important.  The formation, differentiation, modification, disintegration, and reformation of objects is an integral part of psychological growth and development.  For example, the psychologist Erich Neumann used a hierarchical graph structure to represent the development and differentiation of the Great Mother object from the undifferentiated uroboric stage that exists in infancy through to the sophisticated Sophia, Gorgon, Isis, and human objects that occur in adulthood, as shown in Figure \ref{fig:img38}.

\paragraph{Conjecture 2:  Biological Information Processing}

\begin{conj1}
\label{con:inf-proc}
Every biological organism, as a thermodynamic system, is capable of an approximately constant rate of information processing and refinement.
\end{conj1}
\vspace{4pt}

It is important to distinguish the type of information processing performed by a person from that performed by a computer.  The information processing performed by a biological organism is a function of the inputs and outputs of the objects $A$ in equation \ref{eqn:cie}.
There is an interesting connection between Conjecture \ref{con:inf-proc} and the Big Five personality traits - openness, conscientiousness, extroversion, agreeableness, and neuroticism.   The metric of openness, which is correlated to creativity, intelligence, and novelty-seeking behavior, may be understood thermodynamically as a measurement of an individual's rate of internal content generation and information processing.  This refers to the capacity of the individual to:
\begin{enumerate}
    \item internalize information from his environment,
    \item synthesize this information internally,
    \item generate new internal content, and
    \item modify his environment through the externalization of inner content by outward acts of creation.  This item also increases the CIE of the environment and generally improves the environment for others.
\end{enumerate}

When combined with Conjecture \ref{con:self-destruct}, Conjecture 
\ref{con:inf-proc} leads to the following corollary.

\paragraph{Corollary 2.1:  Counter-Processing}

\vspace{4pt}
\begin{corollary}
\label{cor:cor1}
If an individual, as a thermodynamic system, is capable of a constant output of information processing and refinement, and this capability is suppressed by the environment, the individual may destructively direct some of his or her processing capabilities inward in order to maintain outward thermodynamic equilibrium.
\end{corollary}
\vspace{4pt}

At the level of the individual, this could appear as neuroticism, dysfunction, or mental illness.  Perhaps unsurprisingly, there is another connection between Corollary \ref{cor:cor1} and the Big Five model.  Corollary \ref{cor:cor1} describes inner turmoil caused by an individual's internal objects attacking each other.  This helps to explain why the trait of neuroticism - meaning the tendency to experience negative emotions, such as anger, anxiety, or depression - is positively correlated with self-awareness.

\subsubsection{Implications for Artificial Intelligence}

The possibility of human-like artificial intelligence, or artificial general intelligence, is an open question in AI research.  If the human psyche can be represented as a series of overlapping and interacting objects at various levels of analysis, then a human-like artificial intelligence would likely need to have a similar object-relations construction.

Jungian psychology suggests there exist archetypal neural structures in the psyche which are common to everyone.  The archetypes may correspond roughly to figures in Greek mythology or other mythological traditions.  Per the Good Regulator theorem, if there exists a general blueprint for the psyche that includes archetypal objects, this blueprint must be an effective representation of the world of human experience.

In order for an AI to be comprehensible to a human, or vice versa, any human-like or general artificial intelligence would likely need to have a similar blueprint.  A parallel to this concept may be found in mythological story of the Titanomachy, a battle in which the human-like Olympians fought and defeated the inhuman Titans.  Titanic characters such as the Hecatoncheires (the hundred-handed ones), the Cyclopses (with single eyes), and the Gorgons (with snakes for hair) suggest remixing of qualia reminiscent of the Dalí painting in Figure \ref{fig:img36}.  A psychoanalytic reading of this myth suggests a stage of human development where the Titans - which are incomplete, malformed, or imperfect as archetypes - are replaced by the Olympians, which are perfect or complete psychic structures.  

\paragraph{Conjecture 3:  Titanic Psychology and AI}

\begin{conj1}
\label{con:titan}
Unless specifically designed to have the same object-relations architecture as the human psyche, any artificial general intelligence is likely to be Titanic, meaning it contains information-processing objects that do not correlate to those found in human experience.
\end{conj1}
\vspace{4pt}

\subsubsection{Implications for Social Systems}

As described in Chapter 1, our economic systems are incentivized to expend as much energy as possible at all times.  In light of the analysis in this chapter, the cause of the need for useless labor becomes clear.  The principle described in Conjecture \ref{con:self-destruct} can apply to social structures in addition to individuals.  At the level of society, examples of self-interference, inefficiency, or wasted energy could appear as bureaucracy, poverty, or the various other modes of dysfunction and suffering listed on page 2.

%% file: chaos05-Conclusions.tex
\subsection{Conclusion}

This chapter has described an object-relations technique as a method to understand and model general types of complex systems.  A mathematical description of the amount of information contained in complex systems is provided in the form of the Complex Information Entropy (CIE) equation.  Additionally, a Structure-Learning Algorithm is defined in order to infer the composition of complex systems, and a Zoom-in-Zoom-out (ZIZO) algorithm is provided in order to discover the connections among objects across a range of length- and timescales.  There are potentially far-reaching implications for these methods:
\begin{enumerate}
    \item In the field of ecology, researchers have recognized the species entropy of an ecosystem is correlated to its health, but have also recognized the imperfections in this method and have been searching for more accurate methods to measure environmental complexity.
    \item Comparable problems and applications exist in sociology, in regard to modelling the health of individuals in a community.
    \item In contrast to a capitalist economy, a tropenomy incentivized to optimize CIE could optimize the well-being of its participants without requiring constant growth.
    \item The continuous attractor networks used in neuroscience are an attempt to discover Pareto-optimal mathematical representations, in the form of compact topological spaces, to understand the coordinated activity of individual neurons relative to higher-order neural functions \cite{gardner2022toroidal}.
    \item In psychology, researchers are lacking in scientific methods to describe cognitive processes below surface-level observation.  There have been attempts in recent years to explain cognitive processes using theories based on entropy and thermodynamics, such as the Bayesian brain theory and the free energy principle, but these models assume the internal unity of the systems under investigation and do not account for modes of inner conflict, faulty reasoning, or emergence.
    \item In the field of biology, the CIE equation resolves the dark room paradox, which arises if one assumes an organism's nervous system is incentivized to minimize the entropy its predictive model of the world.  The organism could bring its predictive error to zero by sitting in a dark room and not observing anything.  The paradox is removed if one considers an organism's highest goal is to maximize CIE rather than to minimize prediction error, as this is equivalent to the organism developing the richest and most complete understanding possible of its environment, while maximizing $C$ includes minimizing object entropy as a subcomponent.
\end{enumerate}

Other potentially groundbreaking applications exist in the fields of engineering design, the physical sciences, and medicine.  As mentioned in Section \ref{section:ml}, if reinforcement learning is applied to a ZIZO-type algorithm, it creates the ability to design systems across multiple length- and timescales.  This creates the potential to generate complex engineering systems with an effectiveness beyond the capability of human experts.  Other possible applications exist in chemistry and materials science.  The ability to correlate molecular structure to macroscale chemical and material properties, and to design chemicals and materials with desired properties, is a major unsolved problem.

Similarly, the possibility exists to repair cells and tissues by using RL to discover cellular mechanisms of action, which are often unknown to medical researchers.  The development of treatments for medical conditions is difficult and time-consuming due to complex interactions present in the body.  The process often involves years of laboratory research and clinical testing.  In the long-term, the possibility exists to destroy individual cancer cells, rebuild cells in the nervous system, and make other highly specific and individualized cellular interventions.  Combining RL and active learning with the CIE equation creates a method to generate solutions to complex problems that hasn't been available previously.

\subsubsection{Key Takeaways}
\begin{enumerate}
    \item The Complex Information Entropy (CIE) equation is a useful and universally-applicable method to quantify complexity.
    \item The CIE equation explains the behavior of biological organisms.
    \item Measuring the propagation of entropy through a system is an effective way to identify important information at various length- and timescales.
    \item When the environment of a complex system is not well-suited to the system, the system begins to operate self-destructively in order to maintain thermodynamic equilibrium.
\end{enumerate}

%% file: 04-Chapter3.tex
\newpage
\clearpage

\twocolumn[
  \begin{@twocolumnfalse}
    \section{Chapter 3\\The Tropenomy:\\An Economic Organism that Maximizes Complex Information Entropy}
    \vspace{2pt}
  \end{@twocolumnfalse}
]

\PARstart{S}{ystems} with larger CIE are richer or deeper than those with smaller CIE. Intuitively, this is observed in multiple ways across multiple areas of life.  For example, it can be seen in the enjoyment and personal growth that accompanies novel experiences, in environmental ecosystems with a large number and variety of species, and in marketplaces with a diverse array of small businesses.  Having identified the thermodynamic cause of the problems with our society, as well as defining a general method to analyze complex systems of any type, we will now apply this insight to the hypothetical construction of a social structure that is self-correcting, self-improving, and capable of existing in thermodynamic equilibrium.

\vspace{2.0ex plus .5ex minus .2ex}
\subsection{Maxwell's Demon}

Maxwell's demon is a thought experiment that applies to classical entropy.  In the experiment, a demon controls a small door between two closed chambers, each containing a volume of ideal gas.  The demon can open and close the door to allow fast-moving molecules to pass through the door in one direction, and slow-moving molecules in the other direction.  Hypothetically, this would raise the temperature of gas in one chamber while lowering the temperature in the other, violating the second law of thermodynamics that states the entropy of a closed system can only increase.

The second law of thermodynamics is preserved by information theory.  Maxwell's demon consumes a resource, blank memory, to accomplish its task.  The demon must delete information from memory in order to continue running, as it won't be able to operate if its memory is full.  Deleting memory creates heat according to Landauer's principle, which defines the a minimum amount of energy required to erase one bit of information.  Therefore a Maxwell's demon for entropy is theoretically impossible.

Is it possible for a Maxwell's demon to exist for complexity rather than entropy?

\subsubsection{Conjecture:  Maxwell's Complexity Demon}

\begin{conj1}
    \label{con:maxwell}
    A Maxwell's demon will never be able to maximize CIE in a complex system built to maximize energy usage.
\end{conj1}
\vspace{4pt}

A Maxwell's demon in this context can be understood as a rational agent that is able to modify the objects in the system at will.  As discussed in the previous chapter, such a system will inevitably operate self-destructively to greater or lesser extent, reducing CIE.  The demon may be able to take specific actions to increase CIE, but will be fighting an uphill battle.  Similar to the entropy demon, it's likely not possible for the complexity demon to have the computational resources needed to completely reverse the thermodynamic pressure of a system built to maximize energy output.  Moreover, for such an intelligence to function optimally, it would have to contain a perfect internal model of the complex system itself in addition to the capability to model all possible permutations of the system, which would quickly become computationally intractable.

\subsubsection{The Political Spectrum}

If Conjecture \ref{con:maxwell} is correct, then we have found the underlying principle to unify and subsume the different ends of the political spectrum in regard to economics.

More or less, individuals on the right wing of the political spectrum believe that allowing markets to operate freely without government intervention maximizes the wealth and well-being of a population, while those on the left believe that unregulated markets cause problems, but these problems can be rectified if there are sufficient government programs and regulations in place to act as safety nets.  In effect, the right advocates for the two-regime system from Figure \ref{fig:img1} that attempts to maximally exploit the individuals in the gamma distribution, while the left advocates for a Maxwell's complexity demon to counteract this effect, which is only somewhat possible.

The principle of Maxwell's complexity demon applies to both sides of the political spectrum.  In the same way that a government is unable to fully counteract the negative effects of the incentives that give rise to the two-regime system, an individual can't act as a complexity demon themselves by choosing to financially support only socially-beneficial organizations and activities, as they don't have the necessary knowledge or processing power.

Both political approaches may be partially successful for some fraction of individuals, but neither approach can be optimized for everyone.  For that to occur, the incentive of the social engine must be changed from maximizing energy output to maximizing CIE.  Here we have split the horns of the economic political dilemma and found the Fichtean synthesis of these opposing viewpoints. 

\subsection{Construction of a Tropenomic System}

The solution to the problems listed in Chapter 1 is to rewire the capitalist engine to maximize Complex Information Entropy rather than energy.  As a variation of the word ``economics", stemming from Greek roots meaning the management of a household, this system is termed tropenomics, meaning the management of transformation.  The CIE equation is repeated here for convenience.  


\begin{equation}
    \tag{\ref{eqn:cie}}
    \begin{gathered}
        \textrm{Complex Information Entropy } \mathnormal{}C^*(S) = \\
        \sum_{A_i \in \mathbb{A^*}}c_iH(A_i) + \underset{B_{ij} \in \mathbb{B}^*\mathnormal{}}{\sum_{A_i, A_j \in \mathbb{A^*}, \ \mathnormal{}}}d_{ij}H(B_{ij}(A_i, A_j))
    \end{gathered}
\end{equation}

In a tropenomic system, money is equated directly to Complex Information Entropy.  An activity that increases CIE generates wealth, while an activity that decreases CIE reduces wealth.  Consequently, it is necessary to determine the terms $A$, $B$, $c$, and $d$ in the equation above.

Details of implementation:
\begin{itemize}
    \item Ideally, $A$, $B$, $c$, and $d$ exist in a decentralized model, available to all, and uncontrollable by any single authority.
    \item These models become more detailed, refined, and comprehensive over time.
        \begin{itemize}
            \item There must be a method to improve objects.  Similar to the example of the ML algorithm \ref{alg:deduction} learning an increasingly accurate representation of the ball in Figure \ref{fig:img30}, if a better model can be found for an object, the system should be able to update with the improved model.
        \end{itemize}
    \item There must be a method to add, merge, or dissolve objects.
    \item As discussed in the equation of complexity for turbulence, Equation \ref{eqn:eqn1}, the CIE equation is modular.  Any object in the system only needs to know the interactions between itself and its neighbor objects.
    \begin{itemize}
        \item A large-scale simulation of the entire tropenomy is not needed to calculate individual interactions, which makes the system computationally tractable.
    \end{itemize}
    \item The system is highly democratic, as everyone involved in with any particular object in the model should be able to contribute to the determination of the $A$, $B$, $c$, and $d$ terms relevant to them.
    \begin{itemize}
        \item For example, the next sections discuss environmental restoration and the medical health of individual communities.
        \item Any individual in a particular community should have a say in determining that community's object and affordance models $A$ and $B$ and coefficients $c$ and $d$.
        \item The global population should have a say in determining the relevant terms for the earth's climate.
    \end{itemize}
\end{itemize}

In contrast to an economy, which is an assembly of adversarial agents, a tropenomy can be seen to function more like an organism, with forces tending toward optimizing the well-being of its own internal components, optimizing its relationship to its environment, and continuously improving its own capabilities and complexity.  Like a living organism, and unlike modern capitalism, it is capable of existing in thermodynamic equilibrium without the need for constant growth.

\subsection{A Note about Citizen Science}

An electronic-based monetary system is not very different from the modern financial system:  In the United States, 90\% of currency is entirely digital; worldwide this number is 92\% \cite{forbes}.  A tropenomic system requires no centralized oversight, only a publicly-available model of the functional relationships between tropenomic objects, which may be shared and updated by anyone.  Citizen participation in such a system is realistic, helpful, and immediate.  

Comparable to the virus effect discussed in Chapter 2, localized improvements made by individuals should be able to propagate through the system.

Communities at all levels may determine the forms of Equation \ref{eqn:cie} that work best for them.  Some may borrow useful innovations from each other, allowing for the dissemination of object models.  Different communities may find different solutions and structure their local societies differently, leading to the possibility for preservation or amplification of cultural differences rather than global homogenization.  As discussed in the next section, diversity increases CIE values.

\subsection{Case Study:  The Environment}

\begin{figure}[b]
    \includegraphics[width=\linewidth]{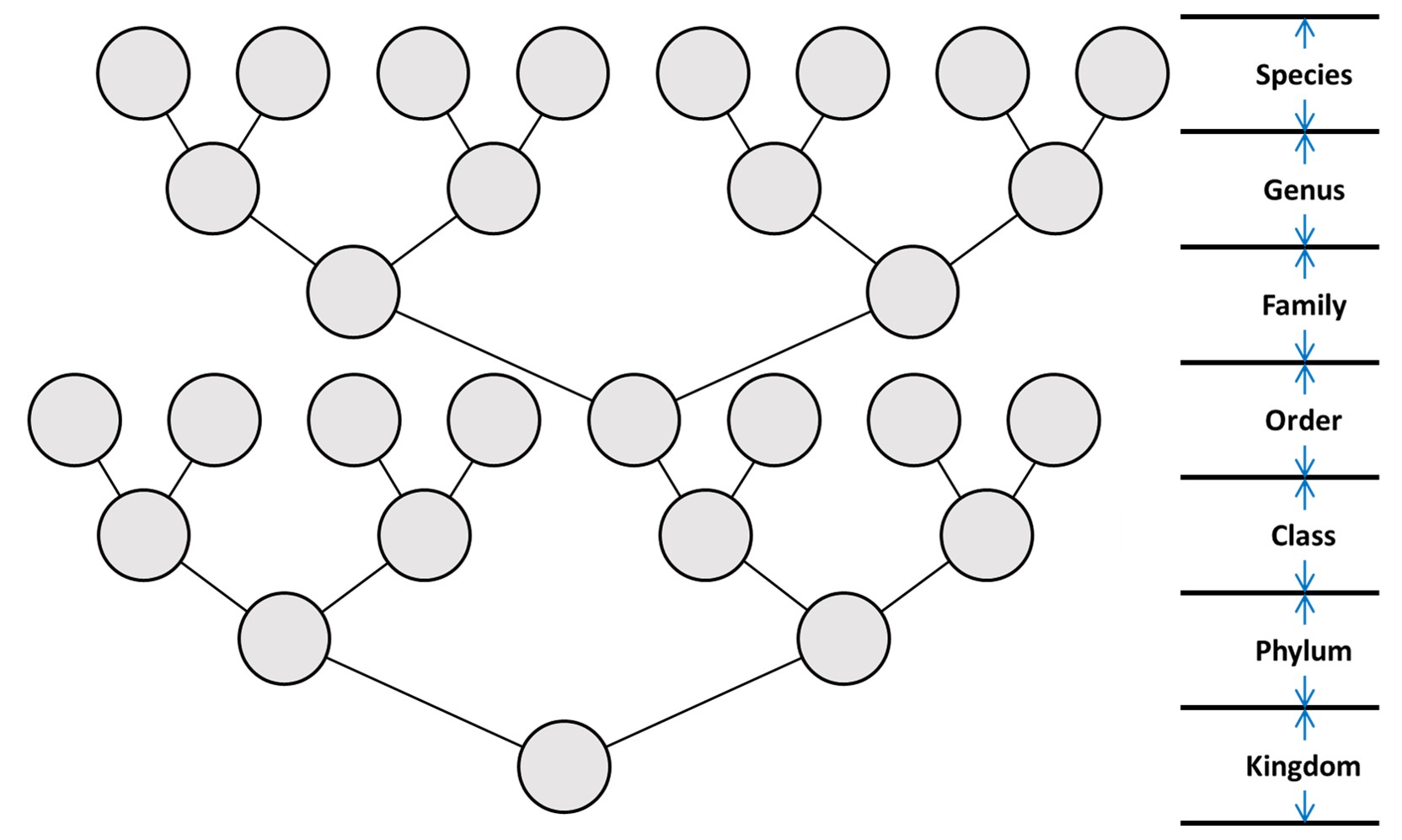}
    \caption{The species classification hierarchy can be used as a tool to quantify CIE in an ecosystem.}
    \label{fig:img52}
\end{figure}

\cite{harte2011maximum}.  In the current state of the ecology community, the diversity of an ecosystem is quantified by diversity $D \approx \mathrm{exp}\mathnormal{}(H(\textrm{population of species}))$, where the entropy $H$ is basically $\sum_i p_i \mathrm{ln}\mathnormal{}(p_i)$ for species $i$, sometimes modified by subjective heuristics such as weighted geometric means.

Researchers have recently included a diversity index $Z$ in this equation in order to capture the intuitive notion that an ecosystem with evolutionarily-diverse species has more diversity than an ecosystem with many species from the same family or genus \cite{leinster2012measuring}. $Z_{ij}$ represents the similarity between species $i$ and $j$.  However, there is currently no scientifically robust way to measure $Z$.

The ecosystem CIE is a more accurate measurement of environmental richness than the diversity metric $D$ or the diversity index $Z$.  A first estimate of CIE can be made by ignoring the interactions among species, corresponding to $B_{ij}$ in equation \ref{eqn:cie}, and considering only the genetic diversity of the population.

\begin{equation}
    \label{eqn:env}
    C = \sum_{i = \textrm{taxonomy level}\mathnormal{}} 2^i\mathrm{ln}(N_i)
\end{equation}

$i$ corresponds to the layer in the taxonomy hierarchy, from $i=1$ at the species level to $i=7$ at the kingdom level, as shown in figure \ref{fig:img52}.  $N_i$ is the number of different entries at that level in the ecosystem.  This system quantifies the notion that adding a species from a different kingdom to an ecosystem increases its diversity much more than adding a new species from an already-present genus.

\subsection{Case Study:  Community Health}

In order for an organism to thrive, it must exist in a supportive environment.  On a physiological level, this includes physical necessities such as air, water, food, heat, and shelter, that are applicable to all organisms from the most simple to the most complex.  More complex organisms, particularly human beings, have higher-order needs that include safety, social belonging, self-esteem, and self-actualization.

The relationship between an individual and his or her environment is an important component of psychological health.  The objects needed to study mental health lie both within the individual and in his environment.  This implies if a model can be developed that includes inner variables within the individual as well as outer variables in the environment, the model can be used to improve the health of a community.

\begin{figure}[ht!]
    \includegraphics[width=\linewidth]{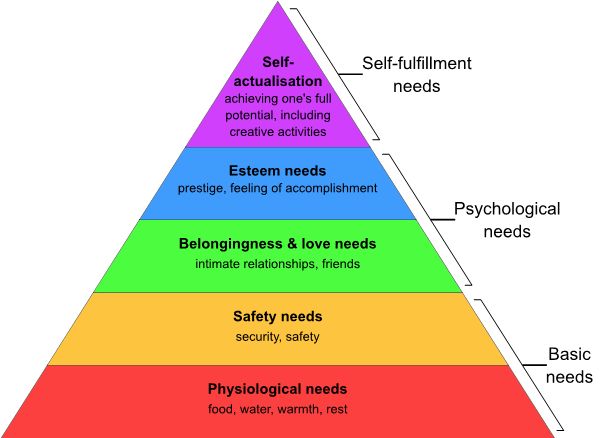}
    \caption{Maslow's hierarchy of needs.  Imaged licensed
under CC BY-SA 4.0 \cite{maslow}.}
    \label{fig:img51}
\end{figure}

An image of Maslow's hierarchy of needs is shown in Figure \ref{fig:img51}.  The bottom layers relate to physical health, while the upper layers relate to psychological health.  According to Maslow's theory, a person must meet the needs on a lower layer before they're able to meet the needs on the layer above.

In order to develop a CIE function for community health, it is reasonable to start with the first two layers of the hierarchy:  physiological needs and safety needs.  Physical needs include air, water, food, heat, and shelter.  Safety needs include physical health, personal security, and financial security.  Similar to the environmental CIE equation for ecosystem health, a community CIE equation for physical well-being could take the form:

\begin{equation}
    \begin{gathered}
        C(\textrm{physical well-being}) = \\ 
        C(\textrm{physiological needs}) + C(\textrm{safety needs}) = \\
        \sum_{A_i \in \mathrm{phys.}}c_iH(A_i) + \sum_{A_i \in \mathrm{safety}}c_iH(A_i) = \\
        \underset{j = \textrm{need}}{\sum_{i = \mathrm{person}}}c_{ij}M_{ij}.
    \end{gathered}
\end{equation}

In this case, $H(A_i)$ is the sum of a Maslow score $M_i$ for each person $i$ in the first two layers of the hierarchy.  More complicated functions can be modelled if desired.  For example, a model could be found that amplifies the negative effect of a few low $M$ scores, which would incentivize making sure no one gets left behind on the lower layers.  As with equation \ref{eqn:env}, coefficients $c$ can be selected to emphasize the importance of lower layers, reflecting the fact that emotional needs can't be addressed before physical needs are secured.

Within a given community, psychological and self-actualization needs are likely to substantially increase if basic needs are met across the population.  

When considering higher-level needs - which include intimacy, trust, acceptance, accomplishment, giving and receiving love and affection, and connections with friends, family, and romantic partners - the best approach is probably not to model an individual's mental state.  Instead, the setup of an ideal system would be such to allow and support each individual to pursue what is meaningful to him or her.

%% file: 05-Chapter4.tex
\twocolumn[
  \begin{@twocolumnfalse}
    \section{Chapter 4\\How to Solve Climate Change\\Building a Digital Twin with an Entropy-Based Fitness
Function}
    \vspace{2pt}
  \end{@twocolumnfalse}
]

\PARstart{I}{f} one imagines the various systems in the biosphere to be represented as a graph of nodes and edges – where the nodes represent ecological processes and edges represent the interactions among these processes – then our economic reward structure tacitly declares the information bandwidth of most of these edges to be zero.  Because economic incentives are the method by which energy and resources are regulated in society, setting these connections to zero results in a condition of runaway energy expenditure.  Reaching an equilibrium of energy balance with the environment is not possible with this incentive structure.  The solution is to develop a mathematical model of our interactions with the ecosystem, called a digital twin, that determines the information bandwidths of the interactions among industrial and ecological processes.  If a such a model is used in combination with a fitness function related to information entropy, then the underlying incentive structure can be changed from one that causes runaway energy use to one that optimizes environmental health.

\vspace{2.0ex plus .5ex minus .2ex}
\subsection{Theory}

Many biological organisms don’t continue to grow throughout their lives.  A human being, for example, grows physically until reaching adulthood, at which point their size and rate of energy usage $E$ become approximately constant.  One can imagine the biological imperative to limit runaway growth, due to the array of problems this would cause. Notably, although physical growth stops, other types of personal growth - including the development of new skills, character improvement, physical fitness, introspection, continuing education, and creative output – can and often do continue.  This distinction is important, because the non-physical modes of growth are related to the Complex Information Entropy $C$ of an individual’s nervous system\cite{casey2022quantifying}.  Depending on the underlying fitness function and reward structures, possible tradeoffs exists between $C$ and $E$ in the design of complex systems.

\subsubsection{Reward Functions}
\label{section:rewards}
\begin{figure}[h]
    \begin{center}
        \includegraphics[width=.9\linewidth]{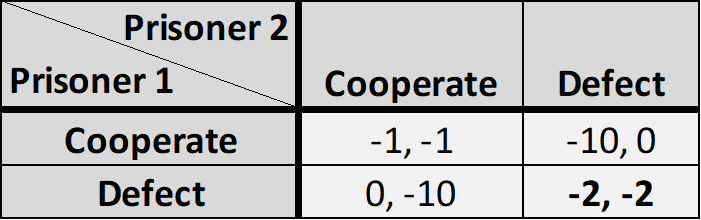}
        \vspace*{2mm}
        \caption{{The Prisoner's Dilemma is an example of a game with a simple reward structure.  A Nash equilibrium exists for the condition where both players defect, as neither player will improve their outcome by changing to another strategy.}}
    \label{fig:img65}        
    \end{center}
\end{figure}

One goal of Game Theory is to study how the reward structure of a game determines the winning strategies and the evolution of player behavior over time. A simple and well-known example is the Prisoner’s Dilemma.  Two prisoners are interrogated in separate rooms, where each can choose to declare the innocence or guilt of the other, and both will receive sentences determined by their combined actions.  Using the reward structure in Figure \ref{fig:img65}, the Nash equilibrium of this game is for both prisoners to defect.  Such a counterintuitive result occurs because each prisoner will spend less time in jail if they choose to defect, regardless of the actions of their partner. 

The Prisoner's Dilemma is a simple example of an important truth discovered by Game Theory:  Depending on how the rewards are configured, even if every player is perfectly rational, it's possible for the best possible strategy to not produce the best possible outcome.

Another example of the influence of reward structure is the famous Duverger’s Law in political science.  This law states that simple-majority, winner-take-all elections, as are commonly held in the United States, will tend to bring about exactly two political parties that are approximately equally matched\cite{duverger1959political}\cite{palfrey1988mathematical}.  This result makes intuitive sense from a Game Theory perspective; multitudinous parties can strengthen themselves through consolidation, while the introduction of a new party will weaken its political allies.  Every non-equilibrium scenario creates pressure toward the stable two-party equilibrium.

\subsubsection{Economic Reward Functions}

For most social institutions, money is used as the prototypical fitness indicator and reward function.  In fact, this observation is ubiquitous enough that it can escape casual notice, similar to the proverbial fish that is unaware of water.   Although it can seem strange to consider dollars to be equivalent to fitness points, this is a common practice.  Measurements like Gross Domestic Product (GDP), the Dow-Jones Industrial Average, and the NASDAQ Composite Index are common fitness functions that are used to gauge social health and well-being.

Implicit in every complex system is a mechanism to distribute energy.  At its most basic level, money functions as a repository for energy via the ability to be exchanged for goods or services.  This role as an energy source explains why money operates as a fitness indicator and reward function.

\begin{figure}[h]
    \includegraphics[width=\linewidth]{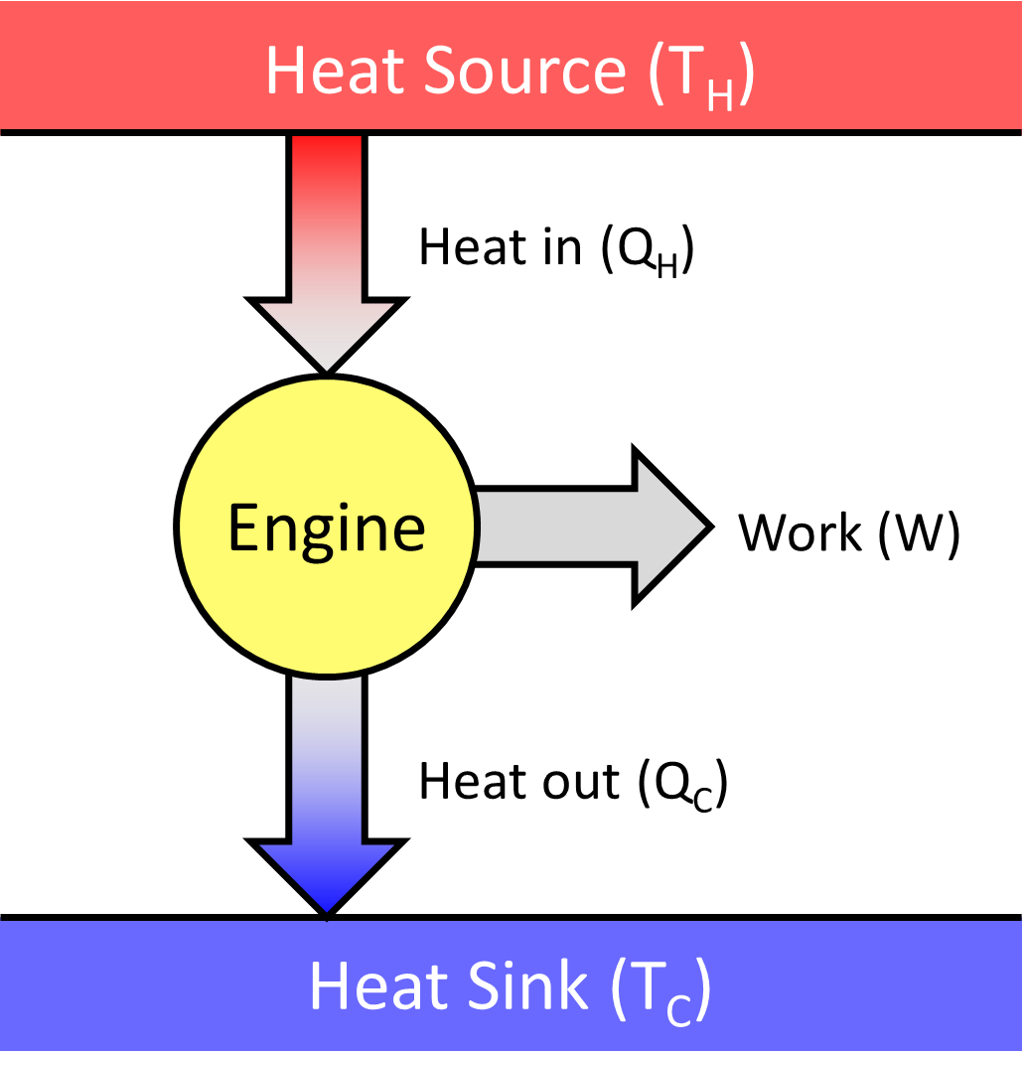}
        \caption{In the most general sense, a heat engine is an object that converts energy into useful work.  Machines, living creatures, and social institutions fit this description.}
    \label{fig:img64}
\end{figure}

Considerations of energy flow and energy balance are universally important to every scientific field.  In regard to biology, each living creature functions as a heat engine.  In the most general description, a heat engine receives energy from an external source, converts some percentage of this energy into useful work, and exhausts the remaining energy to the environment, as shown in Figure \ref{fig:img64}. 

As discussed in Section \ref{section:rewards}, a complex system's equilibrium mode of operation is shaped by its reward function.  Consequently, optimization of the economic fitness function, which is achieved by engaging in profitable activity, determines where and how energy and resources are distributed across the world.  In this particular sense, the economy can be thought to function as the nervous system of our social organism.

\subsubsection{Incorrect Calibration of Rewards}
\label{section:calibration}

One can immediately see that there are problems with the game-theoretic optimum solutions produced by economic reward structures.  Some harmful actions are profitable, while some helpful actions are unprofitable.  For example, there isn't a strong profit motive to find an effective and reliable solution to remove carbon dioxide from of the atmosphere, despite the good this would do.  Conversely, it can be profitable to produce large quantities of junk mail, despite its wastefulness of natural resources.  

\begin{figure}[b]
    \includegraphics[width=\linewidth]{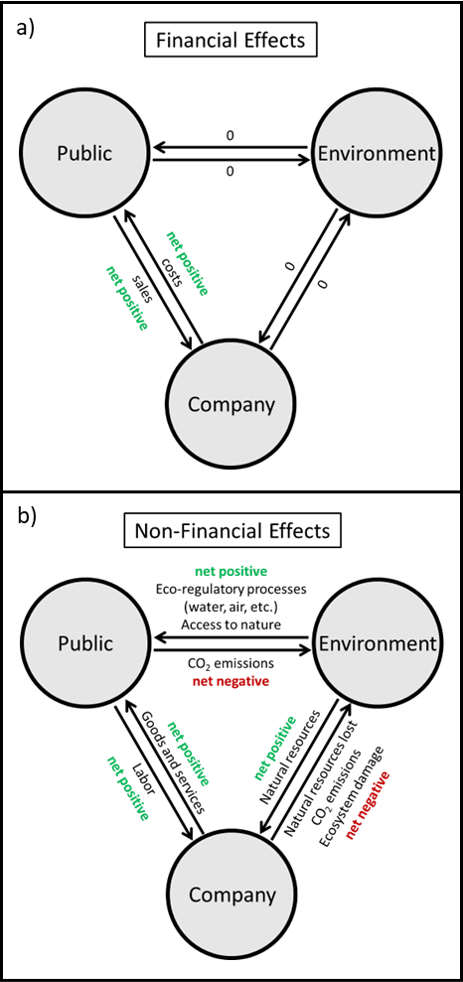}
        \caption{a) The financial system does not quantify the effects between the environment and the general public, or between the environment and industry.  Through omission, these values are tacitly set to zero, which is inaccurate.  b) An accurate fitness function must account for the rich connections between the environment and social institutions.  The directed edges in the graph are equivalent to the communication channels in a digital twin mathematical model.}
    \label{fig:img63}
\end{figure}

The mathematical reason for this miscalibration is shown in Figure \ref{fig:img63}.

The notation $f = {\uparrow} (x)$ is used here to denote a monotonically-increasing function of x, meaning $f(x) > f(y)$ for all $x > y$.  The hypothetical company in Figure \ref{fig:img63} is incentivized to maximize its profit $P$.  $P$ is therefore the company's fitness function.  Maximizing $P$ involves maximizing output of goods and services, which involves maximizing the amount of work performed $W$, which involves maximizing total energy use $E$ assuming the company works efficiently.  Therefore $P = {\uparrow}(W)$, $W = {\uparrow}(E)$, and by transitivity $P = {\uparrow}(E)$.  Because $P$ increases monotonically with $E$, maximizing fitness $P$ involves maximizing energy use.  This is fundamentally why global warming is unstoppable using the reward structure shown in Figure \ref{fig:img63}a.

\subsubsection{The Generalized Maxwell's Demon}

Ideally, one would want the fully-connected system in Figure \ref{fig:img63}b to tend toward a steady-state of energy use, while also optimizing the fitness of the system components.  As shown, setting some of the connections in the graph to zero creates a condition of runaway energy expenditure without regard to environmental damage.  This model is limited in its ability to optimize the actual fitness of social institutions and the environment.  Considering the example of recycling, the reward structure in Figure \ref{fig:img63}a only incentivizes the use of recycled materials when natural resources become depleted enough that recycled materials are less expensive to produce than non-recycled materials.

It makes sense to consider government intervention as a possible solution to this problem.  EPA regulations, carbon taxes, renewable energy initiatives, and federally-funded R\&D programs are certainly helpful and worthwhile.  However, the important question is as follows:  Is it possible that implementing enough social programs, if they're organized in the right way, will be sufficient to solve the problem of climate change caused by miscalibrated economic incentives?   It seems likely the answer is no.  This result is due to the fundamental thermodynamics inherent to information processing.

The example of carbon scrubbing is referenced in Section \ref{section:calibration}.  There isn't a strong financial incentive to develop technologies to remove carbon dioxide from the atmosphere.  The government can therefore decide to encourage production of this technology by funding research and development in this area.  In such a scenario, the government is acting as a rational and intelligent agent, which uses its intelligence to identify problems, to perform reasoning about their causes and effects, and to chose when and how to act.

\begin{figure}[h]
    \includegraphics[width=\linewidth]{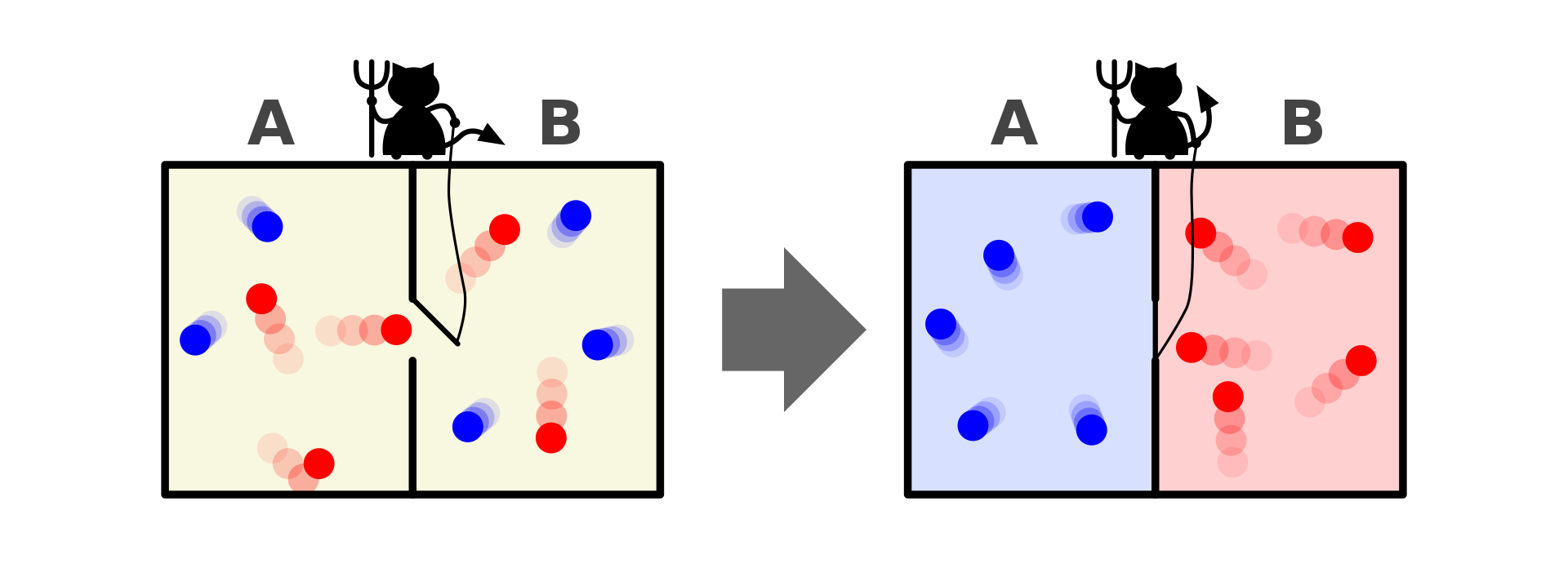}
        \caption{The classical Maxwell's Demon attempts to separate warm air into adjacent chambers of hot and cold air by intelligently operating a doorway between the chambers.\cite{md}}
    \label{fig:img66}
\end{figure}

In this role, such an intelligence is reminiscent of Maxwell's Demon, which is a thought experiment about the possibility or impossibility of an intelligent agent's ability to reverse the second law of thermodynamics.  As shown in Figure \ref{fig:img66}.  The traditional Maxwell's Demon experiment imagines the demon controlling a small door between two chambers containing a volume of ideal gas.  The demon can open and close the door at will, in order to allow fast-moving molecules to pass through the door in one direction and slow-moving molecules to pass in the other direction.  Over time, separating the high-energy and low-energy particles would raise the temperature of gas in one chamber while lowering the temperature in the other.  This violates the second law of thermodynamics, which states the entropy of a closed system can only increase, and must evolve toward thermodynamic equilibrium over time.  

In the most general sense, the ideal gas system experiences pressure to evolve toward equilibrium, while the demon pushes against this pressure by using intelligence and computation.  However, it's physically impossible for the demon to succeed, due to the heat generated by the act of computation as defined by the Landauer limit.  The demon may be partially successful by slowing the system's rate of entropy increase, but is fundamentally unable to reverse the process.

This paper proposes the concept of a Generalized Maxwell's Demon (GMD), which broadens the definition of the classical Maxwell's Demon to include game-theoretic scenarios and other complex systems governed by incentives.  The second law of thermodynamics can be described as a particular example of this type of system, wherein the physics of the system comprise the underlying rules of the game and the system entropy acts as the fitness function to be maximized.  A GMD is defined as a rational agent working to counteract the progression of a system toward the natural equilibrium determined by the system's underlying reward structure.  As with the classical Maxwell's Demon, it's reasonable to hypothesize that no GMD can be more than partially successful.

Returning to the example of $\mathrm{CO}_2$ capture, if the government enables this technology by directing energy toward research and development, then they are acting as a Generalized Maxwell's Demon.  Although funding this type research is a good idea, an accurately-calibrated system wouldn't require the explicit creation of social programs, and instead would naturally incentivize this behavior through its fitness function.

\subsection{Remediation}

Once the problem of misalibrated incentives has been identified, the next question is how to calibrate them correctly.  A mathematical model and a fitness function are needed.  Inspiration can be drawn from NASA's development of Digital Twin technology\cite{piascik2010technology}.

\subsubsection{Digital Twin}
\label{section:dt}

A digital twin is a high-fidelity model of a system used to emulate the real system in as much detail as possible.  For example, when designing a new aircraft, aerospace engineers may create a digital twin of the vehicle to be used in simulated wind tunnel tests\cite{digitalengineering}.  Real-world wind tunnel testing is expensive and time-consuming, and realistically limited to only a few tests per day.  By contrast, thousands of simulated tests can be performed in the time needed to perform a single real-world test.  Physical testing can then be used for verification rather than data gathering:  the more accurately the simulated test data matches the real-world test data, the more confidence can be placed in the accuracy of the digital twin.

The effects of carbon dioxide emissions are a rich ground for mathematical modelling and simulation due to the cascade of complex ecological processes initiated by raising the atmospheric temperature.  One such cascade results from melting polar ice caps, which cause ocean water levels to rise, which causes damage to coastlines and will eventually require a large percentage of the population to migrate inland.  Another cascade results from increasing the average temperatures of terrestrial and aquatic ecosystems, which stresses the survivability of various wildlife species, which lowers the diversity of these ecosystems, which in turn reduces the ability of the ecosystem to absorb and metabolize further environmental stressors.  The result is a continuous cycle of ecosystem damage.

The examples in the previous paragraph span multiple scientific disciplines, including geography, biology, and sociology.  Additionally, the effects caused by climate change operate on every scale ranging from microorganisms to global weather patterns.  A digital twin of the Earth's climate is needed to understand and characterize these effects.

\paragraph{Software Development Tasks}
\label{section:tasks}

A few key software tools are needed to enable the construction this type of model.  An effective digital twin must be scalable, computationally-efficient, Pareto-optimal, multiphysical, and self-improving.  It can be helpful to describe each of these terms individually.  
\begin{enumerate}
    \item Scalable models discover the dominant physics at various lengthscales and timescales.  This often involves the hierarchical inference of physical models from real-world data. Until recently, the inference of hierarchical model structures was primarily done using probabilistic Bayesian methods \cite{margaritis2003learning} \cite{fine1998hierarchical}.  Within the past few years, a research group at MIT has made great strides in developing AI methods to discover physics equations from data \cite{udrescu2020ai}\cite{wu2019toward}\cite{udrescu2020ai2}\cite{liu2021machine}\cite{tegmark2019latent}\cite{liu2022machine}\cite{liu2022ai}\cite{Udrescu_2021}\cite{wu2018}.  Further improvements to these methods are described in Reference \cite{casey2022quantifying}.
    \item Computational efficiency in this context involves dynamic course-graining and fine-graining.  Computation speeds can be improved by not performing simulations in more detail than necessary.  Simulation efficiency can also be increased by using machine-learning methods.  Early results using machine learning are encouraging; when trained to perform fluid dynamics simulations, Google DeepMind's AI is able to produce accurate results in a fraction of the time required for traditional simulation methods \cite{Sanchez2002learning}.  Interestingly, the video game company Ubisoft has also pioneered several fast-physics-learning techniques to improve player experience of in-game physics \cite{holden2019subspace}.
    \item Pareto-optimal models minimize information entropy while maximizing predictive accuracy.  Intuitively, a Pareto-optimal model is a concise description of an important truth, such as $F=ma$.  Pareto-optimality is discussed in detail in References \cite{udrescu2020ai2}, \cite{Udrescu_2021}, \cite{Shea_2021}, and \cite{casey2022quantifying}.
    \item Multiphysical models discover the relationships among variables across different scientific fields.  A multiphysical model of the processes described in the second paragraph of Section \ref{section:dt} would include variables related to the fields of geography, biology, and sociology, among others.
    \item Self-improving models iteratively apply revisions over time.  The iteration process is described in Reference \cite{casey2022quantifying}.  Several types of algorithms can be developed to facilitate continuous improvement, including crawler algorithms, annealer algorithms, grow/shrink algorithms, and reinforcement learning-based theorycrafter algorithms.
\end{enumerate}

\subsubsection{Fitness Function}

In addition to the digital twin, it's necessary to define a fitness function in order to characterize system health.  This leads to the question of how to quantify fitness without resorting to subjective heuristics.  There is an elegant and surprising answer to this question.  The fitness of a complex system appears to be directly connected to its information content.  Specifically, the health and fitness of ecological, sociological, and economic systems seems to increase as their Complex Information Entropy (CIE) increases, which is the amount of information needed to Pareto-optimally describe the system \cite{casey2022quantifying}.  Along these lines, environmental researchers have begun to describe the diversity of species in an ecosystem as a measurement of its information entropy \cite{harte2011maximum}\cite{leinster2012measuring}.

Using this method, it's possible to determine the accurate values for the graph edges in Figure \ref{fig:img63}b.  The strength of these connections is a measurement of the change in complex information entropy caused by the interactions between the objects in the graph, which can be found by referring to the digital twin of the environment and its interaction with social institutions.  The idea of entropy content in a communication channel is reminiscent of signal power in Claude Shannon's seminal paper "Communication in the presence of noise", which describes the increasing power needed to transmit increasing amounts of information in an electronic communication channel \cite{shannon1949communication}.

\subsection{Conclusion}

This chapter has laid the groundwork for a possible solution to the problem of climate change.  Having established the theoretical foundation, the next steps are to perform tests and simulations to find the optimal fitness functions and reward structures, and to develop the mathematical methods and software tools in Section \ref{section:tasks}.  A complete and robust solution for humans to live in equilibrium with the environment appears to be a real possibility.

%% file: 06-Conclusions.tex
\pagebreak
\section{Conclusion}

This work has discussed several ideas, including the analysis of the a capitalist economy as a thermodynamic process, an object-relations method to understand and model complex systems, interactions between a complex system and its environment, a mathematical definition of Complex Information Entropy (CIE), the importance of information content when modelling a system, and the outline of a tropenomic system that uses these ideas for social benefit.  Conjectures 1-4 and Corollary 2.1 are possible avenues for future research.  Additional areas of future work could include the application of CIE models to various types of complex systems, including climate change as discussed in Chapter 4.

A financial system built to maximize CIE requires no centralized oversight, uses a publicly-available model of the functional relationships among tropenomic objects, and can be democratically shared and updated.  There may come a time in the future when today's economy will be seen as a necessary but temporary step on the path toward a system incentivized to maximize the well-being of all participants.  With any luck, topics that today attract significant attention, such as interest rates, inflation, and the Dow Jones Industrial Average, will someday be regarded as artifacts of a primitive past.
\newpage
\clearpage

%% file: main.bbl
\begin{thebibliography}{10}

\bibitem{gamma}
Mark Sweep and C.~Burnett.
\newblock Gamma distribution pdf.
\newblock \url{https://commons.wikimedia.org/w/index.php?curid=10734916}.

\bibitem{powerlaw}
Danvil.
\newblock Probability density function of pareto distribution.
\newblock \url{https://commons.wikimedia.org/w/index.php?curid=31096324}.

\bibitem{wb}
April 2022 global poverty update from the world bank.
\newblock
  \url{https://blogs.worldbank.org/opendata/april-2022-global-poverty-update-world-bank},
  2022.

\bibitem{graeber2018bullshit}
D.~Graeber.
\newblock {\em Bullshit Jobs: A Theory}.
\newblock Penguin Books Limited, 2018.

\bibitem{marx2009economic}
K.~Marx and F.~Engels.
\newblock {\em The Economic and Philosophic Manuscripts of 1844 and the
  Communist Manifesto}.
\newblock Great Books in Philosophy. Prometheus Books, 2009.

\bibitem{locke2002second}
J.~Locke.
\newblock {\em The Second Treatise of Government: And, A Letter Concerning
  Toleration}.
\newblock Dover thrift editions. Dover Publications, 2002.

\bibitem{Chatterjee_2007}
A.~Chatterjee and B.~K. Chakrabarti.
\newblock Kinetic exchange models for income and wealth distributions.
\newblock {\em The European Physical Journal B}, 60(2):135--149, nov 2007.

\bibitem{statista}
Number of amazon.com employees from 2007 to 2021.
\newblock
  \url{https://www.statista.com/statistics/234488/number-of-amazon-employees/},
  2022.

\bibitem{marketcap}
Market capitalization of amazon (amzn).
\newblock \url{https://companiesmarketcap.com/amazon/marketcap/}, 2022.

\bibitem{zip}
Amazon employee salary.
\newblock \url{https://www.ziprecruiter.com/Salaries/Amazon-Employee-Salary},
  2022.

\bibitem{marketwatch}
Amazon ceo jeff bezos total 2020 pay stays at \$1.7 million, with most
  representing security costs.
\newblock
  \url{https://www.marketwatch.com/story/amazon-ceo-jeff-bezos-total-2020-pay-stays-at-1\\7-million-with-most-representing-security-costs\\-2021-04-15},
  2022.

\bibitem{rissanen1978modeling}
Jorma Rissanen.
\newblock Modeling by shortest data description.
\newblock {\em Automatica}, 14(5):465--471, 1978.

\bibitem{kolmogorov1965three}
Andrei~N Kolmogorov.
\newblock Three approaches to the quantitative definition of information.
\newblock {\em Problems of information transmission}, 1(1):1--7, 1965.

\bibitem{solomonoff1964formal1}
Ray~J Solomonoff.
\newblock A formal theory of inductive inference. part i.
\newblock {\em Information and control}, 7(1):1--22, 1964.

\bibitem{solomonoff1964formal2}
Ray~J Solomonoff.
\newblock A formal theory of inductive inference. part ii.
\newblock {\em Information and control}, 7(2):224--254, 1964.

\bibitem{bellman1954theory}
Richard Bellman.
\newblock The theory of dynamic programming.
\newblock {\em Bulletin of the American Mathematical Society}, 60(6):503--515,
  1954.

\bibitem{gray2003language}
Russell~D Gray and Quentin~D Atkinson.
\newblock Language-tree divergence times support the anatolian theory of
  indo-european origin.
\newblock {\em Nature}, 426(6965):435--439, 2003.

\bibitem{wu2019toward}
Tailin Wu and Max Tegmark.
\newblock Toward an artificial intelligence physicist for unsupervised
  learning.
\newblock {\em Physical Review E}, 100(3):033311, 2019.

\bibitem{Udrescu_2019}
Silviu-Marian Udrescu and Max Tegmark.
\newblock Ai feynman: a physics-inspired method for symbolic regression.
\newblock \url{https://arxiv.org/abs/1905.11481}, 2019.

\bibitem{udrescu2020ai2}
Silviu-Marian Udrescu, Andrew Tan, Jiahai Feng, Orisvaldo Neto, Tailin Wu, and
  Max Tegmark.
\newblock Ai feynman 2.0: Pareto-optimal symbolic regression exploiting graph
  modularity.
\newblock {\em Advances in Neural Information Processing Systems},
  33:4860--4871, 2020.

\bibitem{freud1900interpretation}
Sigmund Freud.
\newblock The interpretation of dreams sigmund freud (1900), 1900.

\bibitem{kaufmann1980}
Walter Kaufmann.
\newblock {\em Discovering the Mind. Vol. 3: Freud vs. Adler and Jung}.
\newblock New York: McGraw-Hill, 1980.

\bibitem{Shea_2021}
Daniel Shea and Stephen Casey.
\newblock An information theory approach to physical domain discovery.
\newblock \url{https://arxiv.org/abs/2107.09511}, 2021.

\bibitem{gardner1970fantastic}
Martin Gardner.
\newblock The fantastic combinations of jhon conway's new solitaire game'life.
\newblock {\em Sc. Am.}, 223:20--123, 1970.

\bibitem{glider}
chap06-3.png.
\newblock \url{https://eng.libretexts.org/@api/deki/files/39510/chap06-3.png }.

\bibitem{margaritis2003learning}
Dimitris Margaritis.
\newblock Learning bayesian network model structure from data.
\newblock Technical report, Carnegie-Mellon Univ Pittsburgh Pa School of
  Computer Science, 2003.

\bibitem{rabiner1986introduction}
Lawrence Rabiner and Biinghwang Juang.
\newblock An introduction to hidden markov models.
\newblock {\em ieee assp magazine}, 3(1):4--16, 1986.

\bibitem{turb}
C.~Fukushima and J.~Westerweel.
\newblock False color image of the far field of a submerged turbulent jet.jpg.
\newblock
  \url{https://upload.wikimedia.org/wikipedia/commons/b/b9/False_color_image_of_the_far_field_of_a_submerged_turbulent_jet.jpg}.

\bibitem{coffee}
Ali~Atakan Açıkbaş.
\newblock Close-up shot of a glass of coffee.
\newblock
  \url{https://www.pexels.com/photo/close-up-shot-of-a-glass-of-coffee-5914522/}.

\bibitem{Udrescu_2021}
Silviu-Marian Udrescu and Max Tegmark.
\newblock Symbolic pregression: Discovering physical laws from distorted video.
\newblock {\em Physical Review E}, 103(4), apr 2021.

\bibitem{conant1970every}
Roger~C Conant and W~Ross~Ashby.
\newblock Every good regulator of a system must be a model of that system.
\newblock {\em International journal of systems science}, 1(2):89--97, 1970.

\bibitem{bellman1957}
Richard Bellman.
\newblock {\em Dynamic Programming}.
\newblock Princeton University Press, 1957.

\bibitem{neumann2015great}
Erich Neumann.
\newblock {\em The great mother: An analysis of the archetype}, volume~14.
\newblock Princeton University Press, 2015.

\bibitem{gardner2022toroidal}
Richard~J Gardner, Erik Hermansen, Marius Pachitariu, Yoram Burak, Nils~A Baas,
  Benjamin~A Dunn, May-Britt Moser, and Edvard~I Moser.
\newblock Toroidal topology of population activity in grid cells.
\newblock {\em Nature}, 602(7895):123--128, 2022.

\bibitem{forbes}
Who needs cryptocurrency fedcoin when we already have a national digital
  currency?
\newblock
  \url{https://www.forbes.com/sites/davidblack/2020/03/0\\1/who-needs-cryptocurrency-fedcoin-when-we-already\\-have-a-national-digital-currency/?sh=b577a4d4951e},
  2020.

\bibitem{harte2011maximum}
John Harte.
\newblock {\em Maximum entropy and ecology: a theory of abundance,
  distribution, and energetics}.
\newblock OUP Oxford, 2011.

\bibitem{leinster2012measuring}
Tom Leinster and Christina~A Cobbold.
\newblock Measuring diversity: the importance of species similarity.
\newblock {\em Ecology}, 93(3):477--489, 2012.

\bibitem{maslow}
Androidmarsexpress.
\newblock Simplified hierarchy of needs.
\newblock
  \url{https://upload.wikimedia.org/wikipedia/commons/e/ea/Maslow's_Hierarchy_of_Needs2.svg}.

\bibitem{casey2022quantifying}
Stephen Casey.
\newblock Quantifying complexity: An object-relations approach to complex
  systems.
\newblock {\em arXiv preprint arXiv:2210.12347}, 2022.

\bibitem{duverger1959political}
Maurice Duverger.
\newblock {\em Political parties: Their organization and activity in the modern
  state}.
\newblock Metheun \& Co. Ltd., 1959.

\bibitem{palfrey1988mathematical}
Thomas~R Palfrey.
\newblock A mathematical proof of duverger's law, 1988.

\bibitem{md}
Htkym.
\newblock Maxwell's demon.svg.png.
\newblock
  \url{https://en.wikipedia.org/wiki/Maxwell's_demon#/media/File:Maxwell's_demon.svg}.

\bibitem{piascik2010technology}
R~Piascik, J~Vickers, D~Lowry, S~Scotti, J~Stewart, and A~Calomino.
\newblock Technology area 12: Materials, structures, mechanical systems, and
  manufacturing road map.
\newblock {\em NASA Office of Chief Technologist}, pages 15--88, 2010.

\bibitem{digitalengineering}
Dod digital engineering.
\newblock \url{https://ac.cto.mil/digital_engineering/}.

\bibitem{fine1998hierarchical}
Shai Fine, Yoram Singer, and Naftali Tishby.
\newblock The hierarchical hidden markov model: Analysis and applications.
\newblock {\em Machine learning}, 32(1):41--62, 1998.

\bibitem{udrescu2020ai}
Silviu-Marian Udrescu and Max Tegmark.
\newblock Ai feynman: A physics-inspired method for symbolic regression.
\newblock {\em Science Advances}, 6(16):eaay2631, 2020.

\bibitem{liu2021machine}
Ziming Liu, Bohan Wang, Qi~Meng, Wei Chen, Max Tegmark, and Tie-Yan Liu.
\newblock Machine-learning nonconservative dynamics for new-physics detection.
\newblock {\em Physical Review E}, 104(5):055302, 2021.

\bibitem{tegmark2019latent}
Max Tegmark.
\newblock Latent representations of dynamical systems: When two is better than
  one.
\newblock {\em arXiv preprint arXiv:1902.03364}, 2019.

\bibitem{liu2022machine}
Ziming Liu and Max Tegmark.
\newblock Machine learning hidden symmetries.
\newblock {\em Physical Review Letters}, 128(18):180201, 2022.

\bibitem{liu2022ai}
Ziming Liu, Varun Madhavan, and Max Tegmark.
\newblock Ai poincar$\backslash$'$\{$e$\}$ 2.0: Machine learning conservation
  laws from differential equations.
\newblock {\em arXiv preprint arXiv:2203.12610}, 2022.

\bibitem{wu2018}
Tailin Wu, John Peurifoy, Isaac~L. Chuang, and Max Tegmark.
\newblock Meta-learning autoencoders for few-shot prediction, 2018.

\bibitem{Sanchez2002learning}
Alvaro Sanchez-Gonzalez, Jonathan Godwin, Tobias Pfaff, Rex Ying, Jure
  Leskovec, and Peter~W. Battaglia.
\newblock Learning to simulate complex physics with graph networks, 2020.

\bibitem{holden2019subspace}
Daniel Holden, Bang~Chi Duong, Sayantan Datta, and Derek Nowrouzezahrai.
\newblock Subspace neural physics: Fast data-driven interactive simulation.
\newblock In {\em Proceedings of the 18th annual ACM SIGGRAPH/Eurographics
  Symposium on Computer Animation}, pages 1--12, 2019.

\bibitem{shannon1949communication}
Claude~E Shannon.
\newblock Communication in the presence of noise.
\newblock {\em Proceedings of the IRE}, 37(1):10--21, 1949.

\bibitem{balto-slavic}
Mandrak et~al.
\newblock Indoeuropeantree.svg.
\newblock \url{https://commons.wikimedia.org/w/index.php?curid=5746315}.

\bibitem{kirchhoff2018markov}
Michael Kirchhoff, Thomas Parr, Ensor Palacios, Karl Friston, and Julian
  Kiverstein.
\newblock The markov blankets of life: autonomy, active inference and the free
  energy principle.
\newblock {\em Journal of The royal society interface}, 15(138):20170792, 2018.

\bibitem{Casey2022}
Stephen Casey.
\newblock Improving the engine of society.
\newblock \url{https://zenodo.org/record/7121492#.Y0YEpkzMJPZ}, 2022.

\bibitem{gliders}
img-2.jpg.
\newblock
  \url{https://journals.openedition.org/ejpap/docannexe/image/1640/img-2.jpg}.

\bibitem{baum1966statistical}
Leonard~E Baum and Ted Petrie.
\newblock Statistical inference for probabilistic functions of finite state
  markov chains.
\newblock {\em The annals of mathematical statistics}, 37(6):1554--1563, 1966.

\bibitem{samko2010automatic}
Oksana Samko, A~David Marshall, and Paul~L Rosin.
\newblock Automatic construction of hierarchical hidden markov model structure
  for discovering semantic patterns in motion data.
\newblock In {\em VISAPP (1)}, pages 275--280, 2010.

\bibitem{dagum1991temporal}
Paul Dagum, Adam Galper, and Eric~J Horvitz.
\newblock Temporal probabilistic reasoning: Dynamic network models for
  forecasting. knowledge systems laboratory, medical computer science, 1991.

\bibitem{cigarini2018quantitative}
Anna Cigarini, Juli{\'a}n Vicens, Jordi Duch, Angel S{\'a}nchez, and Josep
  Perell{\'o}.
\newblock Quantitative account of social interactions in a mental health care
  ecosystem: cooperation, trust and collective action.
\newblock {\em Scientific reports}, 8(1):1--9, 2018.

\bibitem{vinuesa2020role}
Ricardo Vinuesa, Hossein Azizpour, Iolanda Leite, Madeline Balaam, Virginia
  Dignum, Sami Domisch, Anna Fell{\"a}nder, Simone~Daniela Langhans, Max
  Tegmark, and Francesco Fuso~Nerini.
\newblock The role of artificial intelligence in achieving the sustainable
  development goals.
\newblock {\em Nature communications}, 11(1):1--10, 2020.

\bibitem{scholten2011every}
Daniel~L Scholten.
\newblock Every good key must be a model of the lock it opens, 2011.

\bibitem{scholten2010primer}
Daniel~L Scholten.
\newblock A primer for conant and ashby’s good-regulator theorem.
\newblock 2010.

\bibitem{negri2017review}
Elisa Negri, Luca Fumagalli, and Marco Macchi.
\newblock A review of the roles of digital twin in cps-based production
  systems.
\newblock {\em Procedia manufacturing}, 11:939--948, 2017.

\end{thebibliography}
